# Assessing Waste Heat Utilization in Power-to-Heat-to-Power Storage Systems for Cost-Effective Building Electrification

Alicia López-Ceballos* and Alejandro Datas*

Instituto de Energía Solar, Universidad Politécnica de Madrid, Avenida Complutense, 30, 28040, Madrid, Spain

* Corresponding author 1: a.l.ceballos@upm.es

* Corresponding author 2: a.datas@upm.es

**Abstract**

Fully electrifying the building sector requires not only the widespread adoption of photovoltaic (PV) self-consumption and heat pumps, but also the integration of cost-effective energy storage solutions. Hybridizing lithium-ion (Li-ion) batteries with power to heat to power storage (PHPS) systems – thermal batteries capable of thermal-to-electric energy conversion – offers a promising and economically viable solution. PHPS systems dispatch combined heat and power by utilizing the low-temperature waste heat generated during the thermal to electric energy conversion process. This study investigates the technoeconomic impacts of waste heat use in PHPS systems integrated with Li-ion batteries and heat pumps to support the decarbonization of the building sector. Two distinct strategies are evaluated: (1) direct use of waste heat to meet heating demands; and (2) the use of waste heat to enhance the heat pump's coefficient of performance. Results show that supplying the waste heat at the demand setpoint temperature is the best solution to integrate PHPS cost-effectively, although enhancing the heat pump's COP with waste heat also yields notable economic gains. Additionally, leveraging waste heat significantly lowers the minimum thermal-to-electric conversion efficiency required for PHPS systems to achieve economic viability. Optimal PHPS designs enable large-scale energy storage and charging capacities, thereby enhancing PV self-consumption rates and reducing the levelized cost of energy. The analysis also reveals that hybridizing PHPS with Li-ion batteries may rise as the optimal solution for moderately priced PHPS systems, with the reduction in levelized cost being more pronounced in solar-dominated regions.

# 1. Introduction

Buildings are responsible for 17% of the global $CO_2$ emissions [1] and only in Europe, they account for 50% of the total heating and cooling demand [2]. As a result, decarbonizing the buildings' sector entails one step ahead towards the net-zero emission target agreed by many countries. Several studies mention the potential of electrifying the heating sector to reduce $CO_2$ emissions [3], [4]. A particularly appealing solution consists of distributed photovoltaic (PV) installations combined with heat pumps [5]. This solution has been shown to achieve lower costs compared to installations relying on PV and natural gas [6], while reaching PV self-consumption ratios of 30-35% [7], [8] or even higher, depending on the heating demand [9].

Incorporating energy storage to the previous configuration, such as thermal [10], [11] and/or electrochemical batteries [12], can contribute to reducing the levelized cost of the consumed energy. Not only does the integration of energy storage systems mitigate mismatch between PV generation and energy consumption, but also it assists in decarbonizing this sector and enhances PV self-consumption to over 60% [10], [12]. As a matter of fact, 80% of the new rooftop solar PV installations in Italy and Germany included batteries in 2023 [13]. In this line, lithium-ion (Li-ion) batteries are widely used in PV self-consumption applications given their high energy density, high conversion efficiency, and declining cost [13], [14]. Due to the widespread adoption of Li-ion batteries, several studies predict a reduction in their cost, lying in the range of 100's \$/kWh between 2030 and 2050 [15], [16]. Despite their expected cost reduction, current optimal self-consumption installations still incorporate low energy capacities of Li-ion batteries.

Currently, no commercially available energy storage technology simultaneously offers both a low cost of energy (~10 €/kWh) and a low cost of power (~100 €/kW). This distinction classifies energy storage systems into: low-energy-cost storage, suited for long-duration applications with large energy but limited power capacity; and high energy cost storage, optimized for short-duration applications with smaller energy but higher power capacity (e.g. Li-ion batteries). To overcome these limitations, recent studies have explored the hybridization of energy storage systems [17]. This solution combines a 'high-power' energy storage, characterized by its high efficiency and high cost of the energy sub-system (i.e., Li-ion battery); with a 'high-energy' storage, characterized by its low efficiency and low energy cost [17], [18]. By leveraging the complementary strengths of both type of storage systems [18], [19] ('high-energy' for baseload and 'high-power' for peak-power demands) the operational flexibility is enhanced. For instance, previous studies have analysed the benefits in power systems of hybridizing a Li-ion battery with Redox Flow Batteries [20], [21], Pumped Hydro Storage [20], hydrogen energy storage [21], [22], [23] or Thermal Energy Storage [24], [25], [26], [27], finding reductions in overall costs, avoidance of oversizing, lower degradation over their lifetime, and enhanced flexibility.

A PHPS system [28], [29], also referred to as Carnot battery [30], [31], [32], [33], electro-thermal energy storage [34], or Thermal Energy Storage [24], [35], is a kind of 'high-energy' storage system that stores electricity in form of heat and convert it back to electricity upon demand. Key components of a PHPS include a Power-to-Heat (P2H) converter for charging, and the Heat-to-Power (H2P) converter for discharging.

P2H systems encompass direct heating methods, such as Joule-based [36], micro-wave [37], or inductive-based converters [38], as well as heat pump-based systems [39]. The latter use ambient air or waste heat as the thermal sources, with such configurations often termed as pumped thermal energy storage [30], [40]. Despite their higher efficiency, heat pump-based systems remain at low technology readiness level and involve significant complexity. Consequently, this work focuses exclusively on PHPS systems employing alternative heating methods than heat pumps, such as Joule-based heating.

H2P conversion may employ well established Brayton and Rankine cycles, both of which have been studied for PHPS applications [30], [31], [41]. Existing PHPS prototypes based on these converters are sized in the MW range [32], [33], achieving efficiencies of 25-40%, consistent with reported H2P efficiencies of these technologies [42]. In contrast, Stirling engines demonstrate high efficiency at smaller scales (~10s kW), making them suitable for self-consumption applications, though their specific cost remains significantly high [42]. More innovative solutions, such as solid-state thermophotovoltaic (TPV) converters, have also been explored for PHPS applications [34], [36], [43], [44], [45], [46], reaching efficiencies over 40% at very small scales [47]. While TPV technology benefits from an inherent modularity, easing their scalability, high efficiencies at larger scales remains unproven in practice.

Because of the relatively low conversion efficiency of H2P converters within PHPS systems (<50%), waste heat recovery may be implemented recirculating heat back into the thermal cycle [29], [48]. Most of these studies focus on PHPS combined with PV transforming heat into electricity through conventional cycles. Alternatively, waste heat is employed to meet the heating demand of industry or buildings. This combined heat and power (CHP) strategy has been assessed to enhance overall round-trip efficiency (RTE), with some configurations achieving 80-90% [30], [33], [36], [49].

The integration of PHPS into residential buildings has been a focus of techno-economic research, often considering the grid as back-up for electricity and a natural gas boiler as back-up for heating. Such studies conclude that PHPS can reduce $CO_2$ emissions and enhance PV self-consumption while maintaining minimum levelized cost of energy [28], [50]. Datas et al. [36] extended the scope to a type of TPV-based PHPS, also referred to as Latent Heat Thermo-Photovoltaic battery, in a fully electrified building through electric heaters, and found it to be competitive with Li-ion batteries in high-energy demand buildings. Our most recent study [49] further explored the hybridization of PHPS with Li-ion batteries and heat pumps in an electrified residential dwelling, demonstrating a robust synergy between both energy storage systems and economic feasibility at PHPS costs consistent with values reported in the literature.

Despite the growing interest in PHPS in recent years, a critical gap remains in the evaluation of waste heat utilization to meet a building's heating demand. To the best of the authors' knowledge, existing studies leveraging waste heat to supply the heating demand take the availability of waste heat for granted and assume it is directly supplied at the required demand setpoint temperature. This assumption overlooks a key operational reality: the temperature of the waste heat generated can vary significantly depending on the H2P technology. For instance, while Brayton engines typically deliver high-grade heat (>100 ºC) [51], Rankine- and Stirling-based technologies provide waste heat at 70-100 ºC [41] and 55-60 ºC [30], respectively. Furthermore, for certain technologies such as TPV cells, a trade-off exists between energy conversion efficiency and waste heat temperature, with efficiency often favouring waste heat temperatures below 60 ºC.

This work addresses these gaps by analysing two strategies for leveraging waste heat from PHPS to meet heating demands, depending on the supply waste heat temperature. First, direct utilization of waste heat when its temperature meets or exceeds the demand setpoint. Second, the use of waste heat to boost the coefficient of performance (COP) of a heat pump, similar to the operation of ground source or solar-assisted heat pumps[52], [53]. Additionally, we extend the analysis beyond dwelling-level [49] to different large building types with high and varied energy demand profiles, where large PHPS units can capitalize on economies of scale to reduce costs [36]. Finally, the climate impact on the PHPS feasibility has been analysed for being a critical factor determining building energy demand.

This article aims to analyse the PHPS economic feasibility addressing the following questions:

- Is there any economic benefit from leveraging the waste heat generated during the heat-to-power (discharging) conversion process of the PHPS?
- What is the impact of PHPS waste heat utilization on the optimal solution?
- How does the discharging efficiency of PHPS, the heat pump's COP, and the waste heat temperature influence the feasibility of hybridizing PHPS with Li-ion batteries?
- Do the results change with different types of buildings and climate conditions?

This study is structured as follows: Section (2) describes the analysed configurations, and the methodology followed to address the questions and the scenarios. Section (3) presents the results and discussion, and Section (4) summarizes the main conclusions.

## 2. System description and Methodology

Three different configurations, depicted in Figure 1, have been studied. In all of them the building is assumed to be fully electrified, meaning that both heating and electricity demand are ultimately supplied by electricity, either from the grid or generated by the PV installation. The base-case configuration, depicted in panel a, comprises a PV installation, a Li-ion battery, a heat pump, and a low temperature energy storage (LTES) system. Two hybrid configurations are presented in panels b and c, both incorporating the PHPS. The Hybrid-60 configuration (panel b) assumes that the waste heat is generated at the temperature required by the application (60 ºC). Therefore, the waste heat is directly stored in the same water tank than in the base-case configuration. On the contrary, the Hybrid-T configuration (panel c) assumes that the waste heat is generated at a lower temperature than the demand ($T$ < 60ºC). In this case, an additional waste heat water tank is included to store the lower-temperature heat at temperature $T$, which is then used to enhance the COP of the heat pump. Further detail on COP calculation depending on the supply temperature is included in Appendix A.

In all configurations, the PV generation surpluses are stored either in the Li-ion battery, the PHPS, or the LTES. Each of the energy storage devices includes two kinds of sub-components: the energy converter(s) and the storage medium itself, which are subject to energy conversion and stand-by losses, respectively. The LTES system consists of a heat pump and a hot water tank. The heat pump's COP is defined as the ratio of the pumped thermal power to the electrical input power ($P_{th}/P_{el}$). An electric heater is included as a back-up for the heat pump when its maximum power capacity is reached. The Li-ion battery comprises an AC/DC converter and the Li-ion cells. The PHPS system is comprised of a Power-to-Heat (P2H) and Heat-to-Power (H2P) converters, and a high temperature energy storage (HTES) unit. The P2H converter is assumed to be a resistive heater and the H2P converter is considered to be a generic heat engine (e.g., Brayton engine, TPV cells, etc.). The HTES includes the high temperature storage material and the thermal insulation.

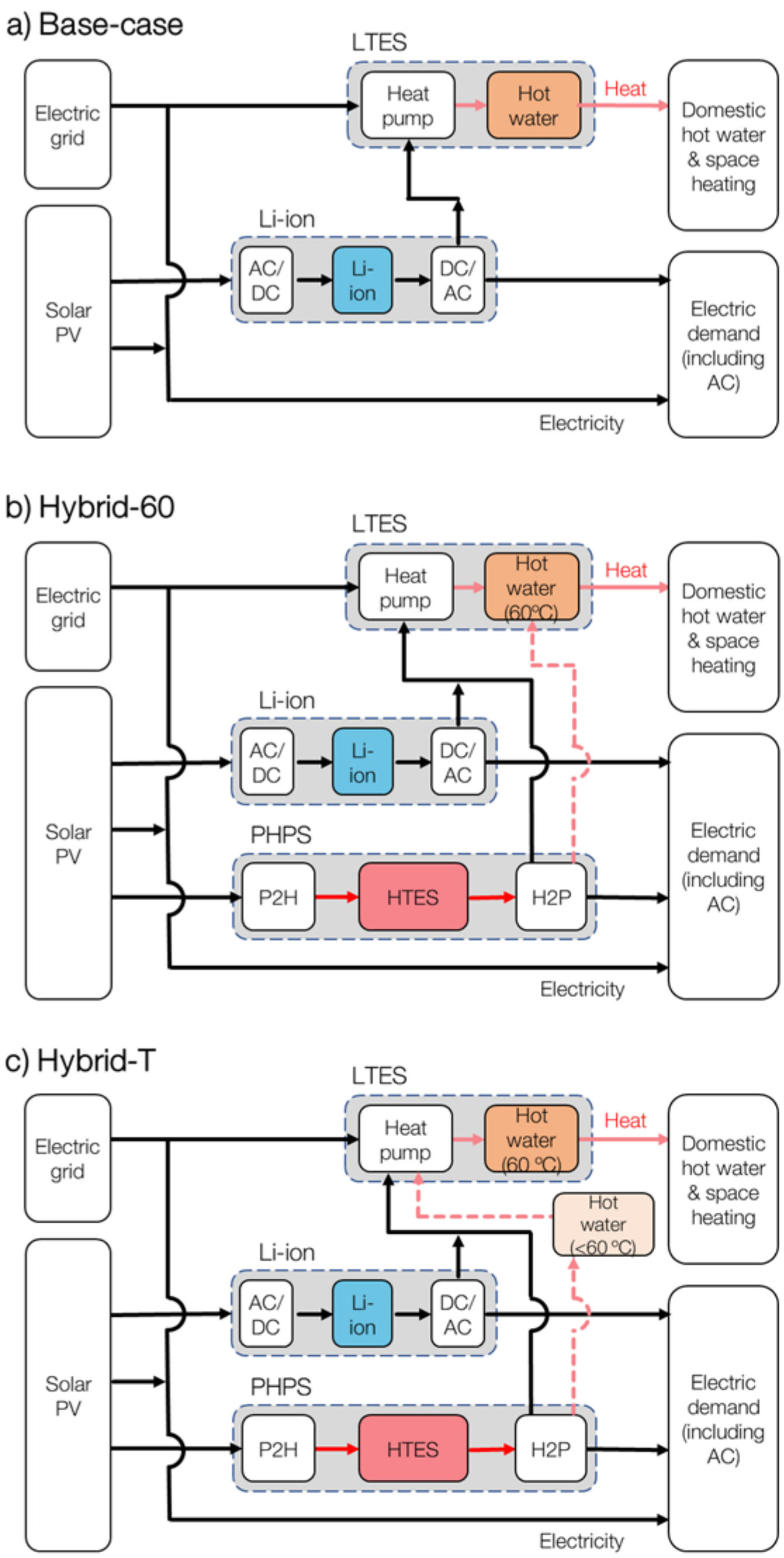

Figure 1. The three system configurations assessed in this work considering only Li-ion battery (Base-case) or hybridizing PHPS and Li-ion battery. Hybrid-60 configuration directly stores the waste heat at 60 ºC in the hot water tank, whilst Hybrid-T stores the waste heat in an additional waste heat water tank at a temperature T<60 ºC and it is used to increase the COP of the heat pump. Black-solid lines represent the flow of electricity, whereas red lines represent the flow of heat, and red-dashed lines represent the flow of waste heat.

A similar model to the one described in [49] has been used to simulate the operation of the base-case and the hybrid configurations. The energy flows within the system are determined to ultimately supply the energy demand (both heat and electricity) of a building at every time step over a year. Energy flows are defined by an energy decision making tree (see Figure A. 1 in Appendix A), based on the state of charge of the energy storage systems, the energy demand, and the PV generation availability at every time-step. The priorities on the energy management are summarized in Table A. 1. The figure of merit is the levelized cost of energy (LCOE), which is minimized for an optimal sizing capacity of each component within the system, which are: the electrical energy and power storage capacities (PHPS and Li-ion batteries); PV power capacity; hot water tank and waste heat water tank energy capacities (optimized separately); and the heat pump electric power capacity. These parameters are indicated in Table 1 (tagged as 'Optimized'), along with a set of techno-economic parameters needed to calculate the LCOE given by [54]:

$$LCOE = \frac{CAPEX + \sum_{t=1}^{T} \frac{OPEX(t)}{(1 + WACC_{nom})^t}}{\sum_{t=1}^{T} \frac{E_{electr}(t) + E_{heat}(t)}{(1 + WACC_{real})^t}} \quad (1)$$

where $CAPEX$ (€) is the initial Capital Expenditure (see Eq. (A.2)), $OPEX(t)$ (€/year) is the yearly Operational Expenditure (see Eq. (A.3)), T is the lifetime of the installation (in years) and $E_{electr}(t)$ and $E_{heat}(t)$ (kWh) are the yearly electricity and heating demand, respectively. Finally, $WACC_{nom}$ and $WACC_{real}$ (see Eq. (A.4)) are the Nominal and Real Weighted Average Capital Cost, respectively. Similarly to previous works [28], [36], [49], [50], the LCOE in Eq. (1) is defined as the total capital and operational costs divided by the total energy demand, including both heat and electricity.[22], [28], [37], [38] The LCOE is calculated from 1-year generation and demand data with 1-hour resolution, assuming every year performs equally. Further detail on the system model, the input data and the optimization procedure are included in Appendices A.1. and A.3.

All the techno-economic parameters of the components described to calculate the LCOE are indicated in Table 1. The cost of each component is detached into energy and power subsystems. The cost per energy capacity of the PHPS ($CPE_{PHPS}$) and the cost per (discharge) power capacity of the PHPS ($CPP_{PHPS\text{-}dis}$) are consistent with typical values found in literature for large scale thermal energy storage systems [34], [36], [38] and heat engines [34], [36], [42]. The CPE and CPP for stationary Li-ion battery are set to 140 €/kWh and 50 €/$kW_{el}$, respectively, corresponding to the forecasted cost in 2030 from ref. [15]. The cost per power of the heat pump ($CPP_{HP}$) has been set to 900 €/$kW_{th}$, also forecasted in 2030 [55]. Given the uncertainty of the lifetime of certain PHPS' elements, all the components in the installation (except for Li-ion) are assumed to last during the PV installation lifetime (25 years). Given that Li-ion's lifetime has been assumed to be 13 years [15], for simplification, we consider one replacement of Li-ion battery as per Eq. (A.2)). Data on the LTES unit have been used for both the hot water tank and the waste heat water tank. The electricity cost corresponds to the fix price for non-household medium size consumers including all taxes in Spain for the second semester of year 2023 available in Eurostat [56]. Simulations in Section 3.2 are conducted for buildings located in Madrid (Spain) and Berlin (Germany) to compare two different climate conditions (PV generation and energy demand). These simulations assume identical grid electricity prices for both locations to neglect its impact on the results. Similar to previous the studies [32], [36], [49], no earnings are assumed from selling electricity surpluses to the grid given the uncertainty and variability of this value in the energy market. Despite there is a trade-off between the cost of the electricity sold and bought from the grid and the cost of the storage systems; the impact of the electricity price variation is out of the scope of this study.

Table 1. Techno-economic model parameters corresponding to large-scale installations.

| Component | Parameter | Value |
|---|---|---|
| PV installation | Cost per power capacity | 500 €/kW [55] |
| | Nominal PV power installed | **Optimized** |
| LTES | Cost per energy capacity | 5 €/kWh [57], [58] |
| | Cost per electrical power capacity | 10 €/kW [57] |
| | Energy capacity | **Optimized*** *The hot water tank and the waste heat water tank are optimized separately |
| | Electric heater power capacity | Set to cover the peak heating demand ($kW_{el}$) |
| | Electric heater efficiency | 100% |
| | Self-discharge heat loss | 0.1 $W·K^{-1}·dm^{-3/2}$ [28] |
| | Temperature of storage | 60 ºC (hot water tank) |
| | | <60 ºC (waste heat water tank, only in Hybrid-T) |
| Li-ion battery | Cost per energy capacity | 140 €/kWh [15] |
| | Energy capacity | **Optimized** (kWh) |
| | Cost per input power capacity | 50 €/$kW_{el}$ [15] |
| | Input power capacity | **Optimized** ($kW_{el}$) |
| | Cost per output power capacity | 0 €/kW |
| | Output power capacity | Equal to the input power capacity |
| | Self-discharge loss | 0 %/day |
| | Round-trip efficiency (RTE) | 90% [36], [55] |
| | Lifetime | 10 y [55] |
| PHPS | Heat to power efficiency | Parameterized in Section 3.1 & 3.2<br>40% in Section 3.2 |
| | Power to heat efficiency | 100% |
| | Cost per energy capacity (CPE) | 15 €/$kWh_{th}$ [34], [36], [38] |
| | Energy capacity | **Optimized** ($kWh_{th}$) |
| | Cost per input power capacity ($CPP_{PHPS-ch}$) | 20 €/$kW_{el}$ [36] |
| | Input power capacity | **Optimized** ($kW_{el}$) |
| | Cost per output power capacity ($CPP_{PHPS-dis}$) | 1000 €/kW [34], [36], [42] |
| | Output power capacity | **Optimized** ($kW_{el}$) |
| | Self-discharge heat loss | 3 %/day [36], [59] |
| | Cost per electric power capacity | 900 €/$kW_{th}$ [55] |
| Electric heat pump | Electric power capacity | **Optimized** ($kW_{el}$) |
| | COP | Parameterized in section 3.1<br>Dependent on waste heat and ambient temperatures in section 3.2 |
| Electricity from the grid | Imported | 0.18 €/kWh [56] |
| | Exported | 0 €/kWh |
| Other economic variables | Nominal weighted average cost of capital | 4% [28] |
| | Inflation | 2% [28] |
| | Lifetime of the rest of the technologies | 25 years [28] |

The implementation of the Hybrid-60 and Hybrid-T configurations is compared with the base-case in 4 different types of building (block of apartments, hospital, hotel and office) in two different locations. To simulate those scenarios, the software Energy Plus® has been used to obtain the hourly heat and electricity demand for each building in its specific location. The PV generation was simulated in PVSyst® for the same location. Further details on the resulting energy demand and PV generation data can be found in Appendix A.

Table 2 below summarizes the yearly energy demand for each building and location obtained from the simulations in Energy Plus. Each building has a different demand profile depending on the day of the week and the season (see Figure A. 3 and Figure A. 4 in Appendix A). The buildings can be categorized based on the heat to total energy demand ratio, being this parameter higher for the block of apartments and the hotel, and lower for the hospital and the office. Our analysis in section 3 focuses on the block of apartments. The same results for a hospital, a hotel and an office are included in Appendix B.

Table 2. Information for each type of building and location. Heat to energy demand ratio is defined as the heat to total energy demand.

| Location | Madrid | | | Berlin | | |
|---|---|---|---|---|---|---|
| Yearly PV generation 1 kW installation | 1.7 MWh | | | 1.1 MWh | | |
| Parameter / Building | Electricity demand (MWh/y) | Heating demand (MWh/y) | Heat to total energy demand (%) | Electricity demand (MWh/y) | Heating demand (MWh/y) | Heat to total energy demand (%) |
| Block of apartments | 580 | 280 | 33 | 560 | 400 | 42 |
| Hospital | 5320 | 710 | 12 | 5200 | 1100 | 17 |
| Hotel | 1400 | 690 | 33 | 1270 | 810 | 39 |
| Office | 380 | 40 | 9 | 370 | 70 | 17 |

# 3. Results and discussions

This section is divided into two subsections in which we have compared the previously mentioned configurations with the base-case. The first section studies the benefits of utilizing of the waste heat when supplied at the setpoint temperature of 60ºC, thus, it focuses on the Hybrid-60 configuration. The second one assumes the Hybrid-T configuration in which the waste heat is supplied at a temperature *T*, lower than the demand setpoint of 60 ºC, thus, it is stored in an additional waste heat tank and used to enhance the COP of the heat pump.

## 3.1. Direct use of the waste heat for cogeneration

Solid-lines in panels a and b of Figure 2 show the relative reduction of the LCOE in the Hybrid-60 with respect to the base-case configurations as a function of the H2P efficiency and the heat pump's Coefficient of Performance (COP). Red-coloured regions indicate cases in which the optimal solution only includes PHPS, whereas blue-coloured regions show layouts in which the optimal solution is the base-case scenario (only integrates a Li-ion battery). Purple-coloured regions depict circumstances where the optimum solution is hybrid (i.e., both PHPS and Li-ion battery are included in the solution). The percentage associated to the colour-scale, φ, represents the share of each storage technology (PHPS or Li-ion battery) on the total capital investment in electrical storage systems of the optimal solution, i.e. $\varphi = \mathrm{CAPEX_{PHPS}}/(\mathrm{CAPEX_{PHPS}} + CAPEX_{Li-ion}) \cdot 100$. Two scenarios are simulated for a block of apartments located in Madrid: one where waste heat is not used (panel a) and one where it is used (panels b). The reader is encouraged to refer to Figure B. 1 in Appendix B for the same results on a hospital, a hotel, and an office. For reference, panels c and d in the same figure depict the minimum LCOE (€/MWh) for the optimal Hybrid-60 configuration without and with waste heat utilization, respectively. It should be noted that the minimum LCOE of the Hybrid-60 configuration, in panels c and d, matches that of the base-case within the blue-regions in panels a and b.

Results in panels a and b in Figure 2 show that, for each COP value, there is a threshold H2P efficiency beyond which PHPS becomes part of the optimal solution, leading to a reduction in LCOE relative to the base-case configuration. This threshold efficiency is significantly lower when waste heat is utilized (panel b), highlighting the relevance of harnessing the waste heat produced by the PHPS system. Panels c and d exhibit a similar trend, where utilizing waste heat results in lower LCOE values.

When the waste heat is not utilized, a high COP is beneficial for the integration of PHPS system, as evidenced by the reduction of LCOE relative to the base case (panel a) and the overall lower LCOE values (panel c). This benefit arises from the reduction in electricity demand at higher COP values, which relaxes the efficiency requirements of the PHPS system. Interestingly, the H2P efficiency threshold and the minimum LCOE (panel c) converge to a relatively stable value and their dependence on COP diminishes as COP increases. This invariance is driven by a significant reduction in the electricity required to meet the same heating demand as COP increases, which limits the additional impact of further COP improvements on the total energy cost. Consequently, beyond a certain COP, further increases in the COP no longer impact on the H2P efficiency threshold.

When the waste heat is utilized, the reduction in LCOE (panel b) is more significant and LCOE is lower (panel d), regardless of the COP or H2P conversion efficiency, thus, promoting the inclusion of PHPS systems in the optimal solution. Remarkably, the inclusion of PHPS in the optimal solution occurs even at H2P conversion efficiencies as low as ~15%. In contrast, when waste heat is not recovered (panel a), H2P efficiencies of at least ~30% are required for PHPS to be considered viable.

Notably, variations in the COP exhibit a minimal impact on the LCOE reduction and on the minimal LCOE when waste heat is utilized (Figure 2, panels b and d respectively). This occurs because waste heat utilization reduces the heat pump's usage to meet heating demand with respect to the case in panel a where waste heat is not used. Furthermore, a higher COP reduces the electricity required to satisfy the heating demand, which further attenuates its impact on overall system performance.

Although a similar trend observed across all four building types, as illustrated in Figure B. 1, hotels and apartments derive greater benefit from waste heat utilization than hospitals and offices. This difference likely stems from the higher relative weight of thermal demand on the overall energy demand. Consistent with this finding, hotels and apartments also exhibit a stronger dependence on the heat pump's COP when waste heat is not utilized, reflecting their greater reliance on heat pumps to supply heat. These results demonstrate that effective waste heat leverage is critical for PHPS systems to lower its minimum discharging H2P efficiency.

It is worth noting that when the H2P efficiency falls below 10%, the PHPS operation is analogous to a Power-to-Heat storage system, overlapping with the functionality of the LTES. Given the lower specific energy cost of LTES (5 €/kWh) than PHPS (15 €/kWh), LTES becomes the optimal choice in such low-efficient scenarios. This economic rationale explains why PHPS is absence in the cost-minimizing solution in our simulations as H2P efficiency is reduced. While a 'zero-efficiency' PHPS configuration could be of interest in space-constrained sites due to its superior energy density compared to water tanks, a detailed evaluation of this niche is outside the scope of this study.

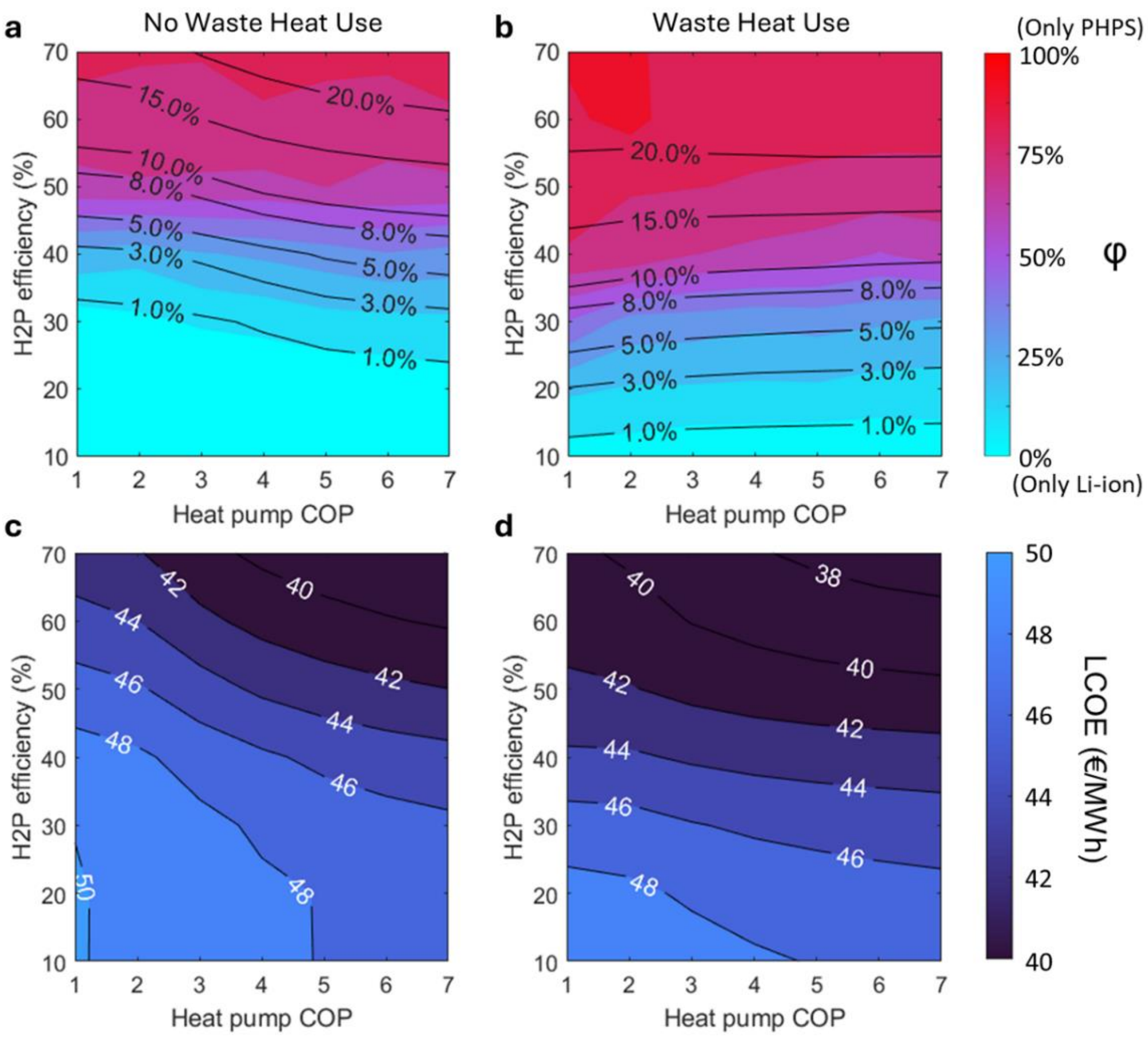

Figure 2. Panels a and b depict the LCOE reduction (black solid lines) for the Hybrid-60 configuration compared to the base-line scenario (only Li-ion battery) for a block of apartments located in Madrid with variations in the COP of the heat pump (x-axis) and the H2P efficiency of the PHPS (y-axis). Panel a shows a scenario in which the waste heat is not leveraged, while panel b illustrates a scenario in which the waste heat is utilized to cover heating demand. Contour plot shows the percentage contribution of each energy storage system within the optimal hybrid configuration defined as $\varphi = \mathrm{CAPEX}_{\mathrm{PHPS}}/(\mathrm{CAPEX}_{\mathrm{PHPS}} + CAPEX_{Li-ion}) \cdot 100$, which is the share of PHPS in the total investment in electrical storage: 100% indicates a configuration relying entirely on PHPS (red-coloured region), whereas 0% represents a system using only Li-ion batteries (blue-coloured region). Panels c and d include the minimum LCOE for the Hybrid-60 configuration for a block of apartments located in Madrid with variations in the COP of the heat pump (x-axis) and the H2P efficiency of the PHPS (y-axis).

Figure 3 shows various parameters for the simulations shown in Figure 2 as a function of H2P efficiency, assuming a COP of 3. Panel a in this figure illustrates the optimal sizing of the three energy storage systems in the solution (PHPS, Li-ion battery and LTES). This panel highlights how PHPS systems replace Li-ion batteries as H2P efficiency increases. Additionally, panel b shows that the discharge time of PHPS (defined as the ratio of energy to total – thermal and electric – output power capacity) is higher than that of the Li-ion battery. The longer discharge time of PHPS is driven by its lower cost of energy storage capacity and its higher cost of output power capacity. For example, at a H2P efficiency of around 30%, PHPS discharge time ranges from 14 hours (without waste heat utilization) to 18 hours (with waste heat utilization). In contrast, the discharge time of Li-ion batteries—when included in the optimal configuration—remains around 5 hours.

When both PHPS and Li-ion batteries are part of the optimal solution that leverages waste heat, their complementary characteristics enable a maximum PV self-consumption of approximately 90% (panel c in Figure 3). The 100% PHPS solution, corresponding to very high H2P efficiencies, results in lower LCOE but a slightly lower PV self-consumption of around 80% (panel c in Figure 3). This underscores the importance of integrating both short- and long-duration storage systems to maximize PV energy utilization. The synergy between PHPS and Li-ion batteries enables effective hybridization, reinforcing the conclusions of our previous study and highlighting their combined potential in enhancing PV self-consumption.

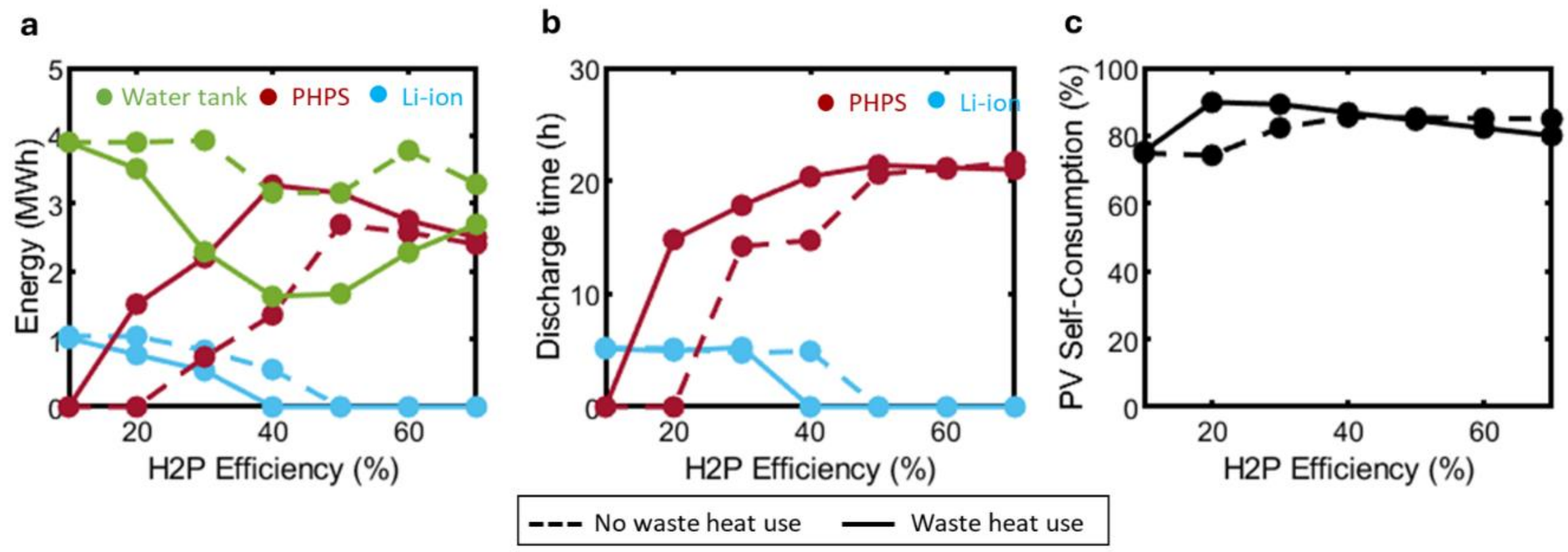

Figure 3. Results of the optimal solution for the apartments scenario when COP=3 and varying the H2P efficiency not using (dashed lines) and using (solid lines) the waste heat. These panels depict: (a) optimal energy capacity of energy storage components (MWh), (b) discharge time (h), and (c) PV self-consumption (%). Panels a and b depict information for the PHPS (red), Li-ion batteries (blue) and water tank (green) in the optimal solution.

## 3.2. Use of the waste heat to enhance coefficient of performance of the heat pump

In the previous section, we assume the waste heat temperature is supplied at the demand setpoint temperature of 60 ºC. In this section we aim to analyse the impact of yielding waste heat with temperature below the setpoint and alternatively use it to enhance the COP of a heat pump. This configuration is defined as Hybrid-T in panel c from Figure 1.

Recovering the waste heat at lower temperatures can be beneficial for certain types of H2P converters, such as the TPV devices. In these systems, waste heat is generated from the cooling of the TPV cells, and their conversion efficiency decreases linearly as the cell temperature increases [60]. Consequently, the waste heat temperature is constrained by the maximum tolerable operating temperature of the cells.

Nevertheless, even at relatively low temperatures (e.g., 30 °C), waste heat can significantly enhance the performance of the heat pump by boosting its COP, thereby improving the overall system efficiency. The efficiency of an air source heat pump is inherently tied to the ambient temperature, creating an inverse relationship with the heating demand of a building. This anticorrelation is clearly demonstrated when comparing the ambient temperatures in Madrid and Berlin, shown in Figure A. 2, with the corresponding heating demand of the different buildings depicted in Figure A. 3 and Figure A. 4. In contrast, when the heat pump utilizes the waste heat of PHPS, it operates at higher and relatively constant COP, independent of both ambient temperature and heating demand. Figure 4 presents the resulting COP obtained from the simulations for both the Hybrid-T configuration (panel a), assuming a waste heat temperature of 40 ºC and a H2P efficiency of 40%, and base-case configuration (panels b). The simulations are performed for a block of apartments located in Madrid considering techno-economic assumptions summarized in Table 1. Panel a illustrates the aforementioned effect for the Hybrid-T configuration, showing the relatively constant and very high COP year-round, in contrast to the more variable COP observed in the base-case in panel b. This behaviour resembles to that of ground source heat pumps, which typically achieve COPs between 2.5 and 6.5 depending on operating conditions [52], [61], [62], [63].

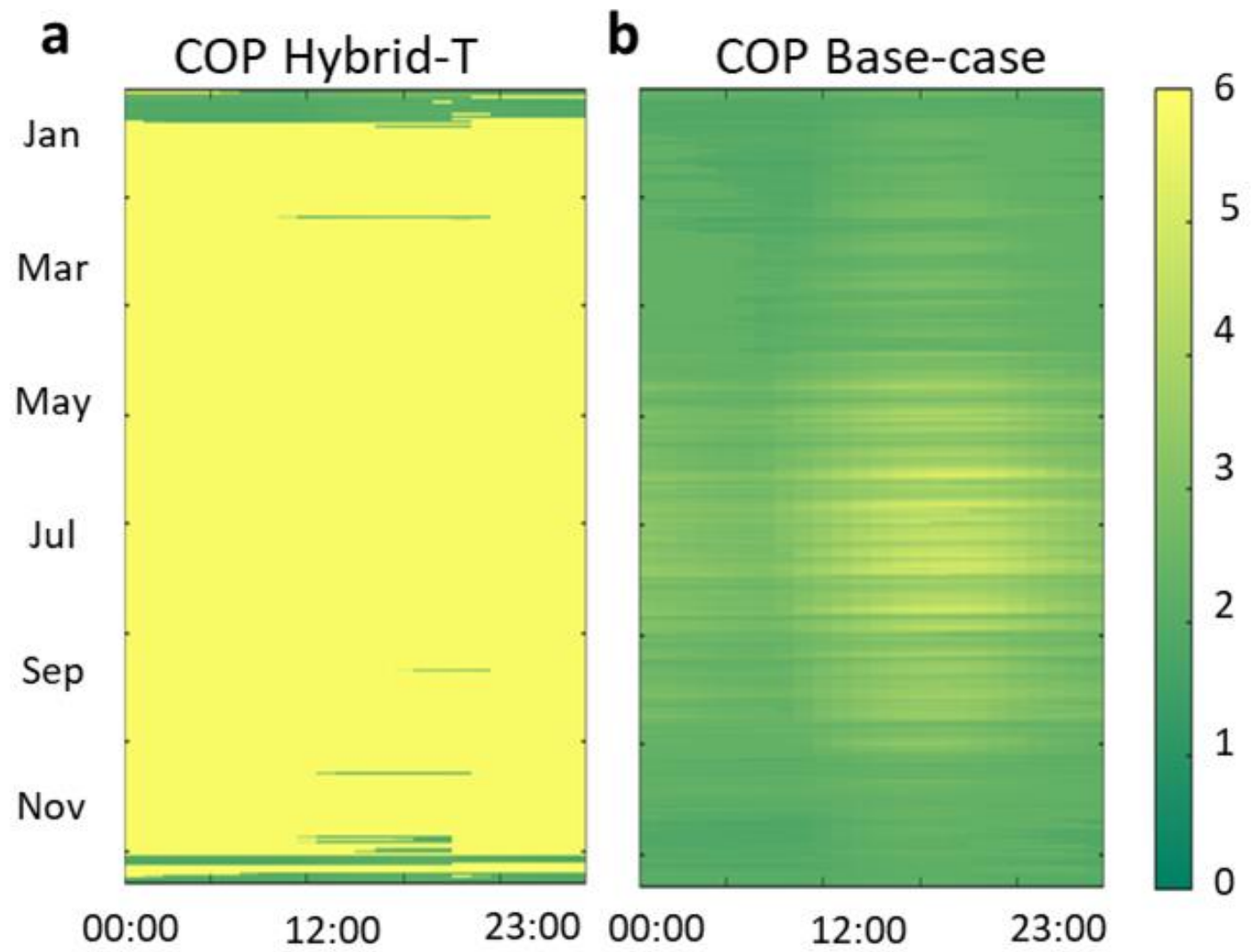

Figure 4. Panels a and b show the resulting COP for each hour (x-axis) and day (y-axis) of the year for the block of apartments scenario in Madrid for the Hybrid-T and base-case configurations respectively, assuming H2P efficiency of 40% and waste heat temperature 40 ºC and techno-economic parameters in Table 1.

To quantify the economic impact of this particular waste heat recovery strategy, panels a and b in Figure 5 show the reduction in the LCOE (solid lines) of the Hybrid-T configuration, relative to the base-case configuration, as a function of the waste heat temperature (from 20 ºC to the demand setpoint temperature of 60 ºC) and the H2P efficiency. Simulations are conducted for a block of apartments located in Madrid (panel a) and Berlin (panel b). The coloured regions indicate the percentage contribution in the capital expenditures of each storage type in the optimal solution. The Hybrid-T configuration excluding the use of waste heat is compared, as a reference, to the base-case beside the panels when the COP of the heat pump depends on the ambient temperature in both configurations. The advantage of using the waste heat at 20 ºC is particularly evident in Berlin, due to its colder climate (ambient temperature<20 ºC for 7500 hours) compared to Madrid (ambient temperature <20 ºC for 6000 hours). For further insights, the reader is encouraged to refer to Figure B. 2, which presents the same analysis for additional building types, including a hospital (panels a and b), a hotel (panels c and d) and an office (panels e and f). For reference, panels c and d in the same figure depict the minimum LCOE (€/MWh) for the optimal Hybrid-T configuration in Madrid and Berlin, respectively.

The results in panels a and b in Figure 5 indicate that the threshold H2P efficiency, above which PHPS becomes part of the optimal solution, decreases as the waste heat temperature increases, reaching a minimum at 60°C. As the waste heat temperature approaches the heating demand setpoint (60 °C), the heat pump's COP increases significantly, theoretically approaching to infinity under ideal conditions. In practice, this reduces the electricity input required to supply the heating demand to negligible levels (for a detailed explanation of COP calculation, refer to Appendix A.1.). Consequently, the Hybrid-T configuration asymptotically converges to the Hybrid-60 configuration, where waste heat is directly supplied at the setpoint temperature of 60 °C. Given this behaviour, the Hybrid-60 configuration consistently emerges as the preferred option.

Interestingly, the integration of PHPS yields a greater relative reduction in LCOE in Madrid than in Berlin if compared to the base-case configuration, and even leads to an approximate halved LCOE in Madrid. Notably, this trend, where PHPS delivers greater benefits in Madrid than in Berlin, holds across all studied buildings, as depicted in Figure B. 2.

Panels a and b in Figure 5 shows that, although PHPS is in principle better suited to Berlin than to Madrid, as evidenced by the greater presence of PHPS in the optimal configuration (larger red regions in panel b than panel a), its associated LCOE in Berlin is approximately twice that of Madrid. A comparable reduction of LCOE in absolute values therefore corresponds to a smaller reduction in relative values. The high LCOE values in Berlin are primarily driven by the lower and less consistent PV availability, which limits the effectiveness of PV, even when combined with PHPS as a long duration energy storage technology, to significantly reduce dependence on the electrical grid. For similar capital expenditures (panel d in Figure B. 5) and total energy demand in Madrid and Berlin (0.87 GWh and 0.96 GWh, respectively), the system in Berlin must purchase a substantial share of electricity from the grid. This results into almost five times higher operational expenditures in Berlin than in Madrid (as seen in panel h in Figure B. 5), and consequent higher absolute LCOE values compare to Madrid. Resulting LCOE values confirm that cost-effectiveness is constrained in northern countries by an inferior solar resource than in southern countries, and that the integration of PHPS partially mitigates this constraint.

Figure B. 4 presents the optimal component capacities for the systems analysed in Figure 5, assuming a discharging (H2P) efficiency of 30%. Both the energy and power capacities of PHPS scale with the waste heat temperature in both locations. However, while PHPS remains part of the optimal solution in Madrid, it is excluded in Berlin at low waste heat temperatures (20 ºC). The lack of PHPS in northern regions may be counterintuitive, as the COP enhancement provided by PHPS should theoretically offer greater advantages in colder climates like Berlin, where lower ambient temperatures and higher heating demands prevail. Nevertheless, the results indicate that the COP enhancement is insufficient to justify PHPS integration in Berlin's optimal energy system under these conditions.

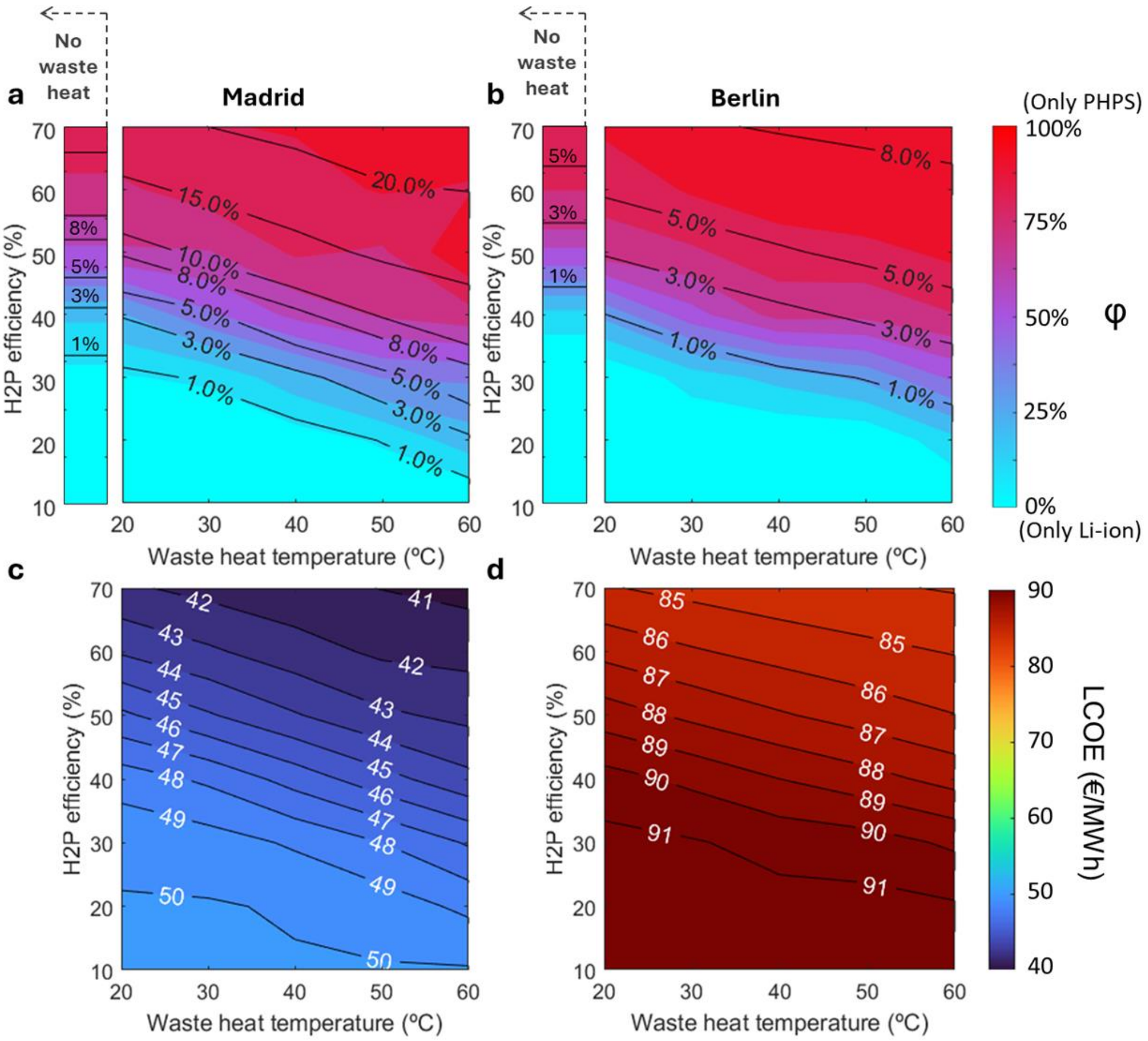

Figure 5. Reduction in the LCOE (solid lines) of the Hybrid-T configuration relative to the base-case configuration for a block of apartments in Madrid (panel a) and Berlin (panel b). Results are provided as a function of the waste heat temperature and the H2P conversion efficiency of the PHPS. Coloured contour plots represent the percentage contribution, in the total capital expenditures, of each energy storage system within the optimal hybrid solution, defined as $\varphi = \mathrm{CAPEX}_{\mathrm{PHPS}}/(\mathrm{CAPEX}_{\mathrm{PHPS}} + CAPEX_{Li-ion}) \cdot 100$, which is the share of PHPS in the total investment in electrical storage: 100% indicates a configuration relying entirely on PHPS (red-

coloured region), whereas 0% represents a system using only Li-ion batteries (blue-coloured region).The case of Hybrid-T in which the waste heat is not used ('*No waste heat*') is also compared to the base case as a reference. Panels c and d include the minimum LCOE for the Hybrid-T configuration for a block of apartments located in Madrid with variations in the waste heat temperature (x-axis) and the H2P efficiency of the PHPS (y-axis).

### 3.3. Sensitivity analysis

Due to all the assumptions study and simplifications of the model, results should be interpreted more qualitatively than quantitatively. Therefore, a sensitivity analysis is performed for the Hybrid-60 configuration applied to a block of apartments in Madrid, assuming a COP of 3, an H2P efficiency of 30%, and a waste heat temperature of 60 °C. The analysis is conducted on the most important parameters within the installation and is included in Figure 6.

Results presented in this figure highlight the CPE of Li-ion and PHPS (panels b and c) as the most influential parameters. Notably, the PHPS system remains feasible with a 50% reduction in the projected 2030 Li-ion's cost. The second most significant factor is the PHPS discharge power cost ($CPP_{PHPS\text{-}dis}$ in panel e), which has a substantial impact when reduced to 100 €/$kW_{el}$. However, the literature suggests that the minimum expected values for this parameter range between 200-300 €/$kW_{el}$ [28], [34]. A reduction in PV costs (panel a) positively influences the inclusion of PHPS within the optimal system, albeit at the expense of reduced Li-ion battery capacity.

Despite increasing the stand-by losses (panel d) beyond 15 %/day hinders PHPS feasibility, reducing these losses have a minimal impact on the overall results. It is important to note the inherent trade-off between PHPS' stand-by losses and its energy cost. Beyond increasing the PHPS energy capacity [36], another strategy to lower PHPS' costs is to reduce thermal insulation investments. Consequently, PHPS developers are encouraged to strike a balance between cost reduction and the associated increase in stand-by losses.

Additionally, setting the electricity imported from the grid to 0 €/kWh (panel i), or excessively increasing the revenue from selling surplus to the grid (panel j) results in the elimination of both storage systems. While keeping these values throughout the year is unrealistic, the expected hourly variations in electricity costs, driven by fluctuations in renewable energy generation over the day, week, or year; could influence the current results.

Variations in the rest of the analysed parameters, the cost per charge power of PHPS (panel f), the nominal weighted average capital cost (panel g) or the heat pump's cost (panel h); exhibit minimal impact on the results.

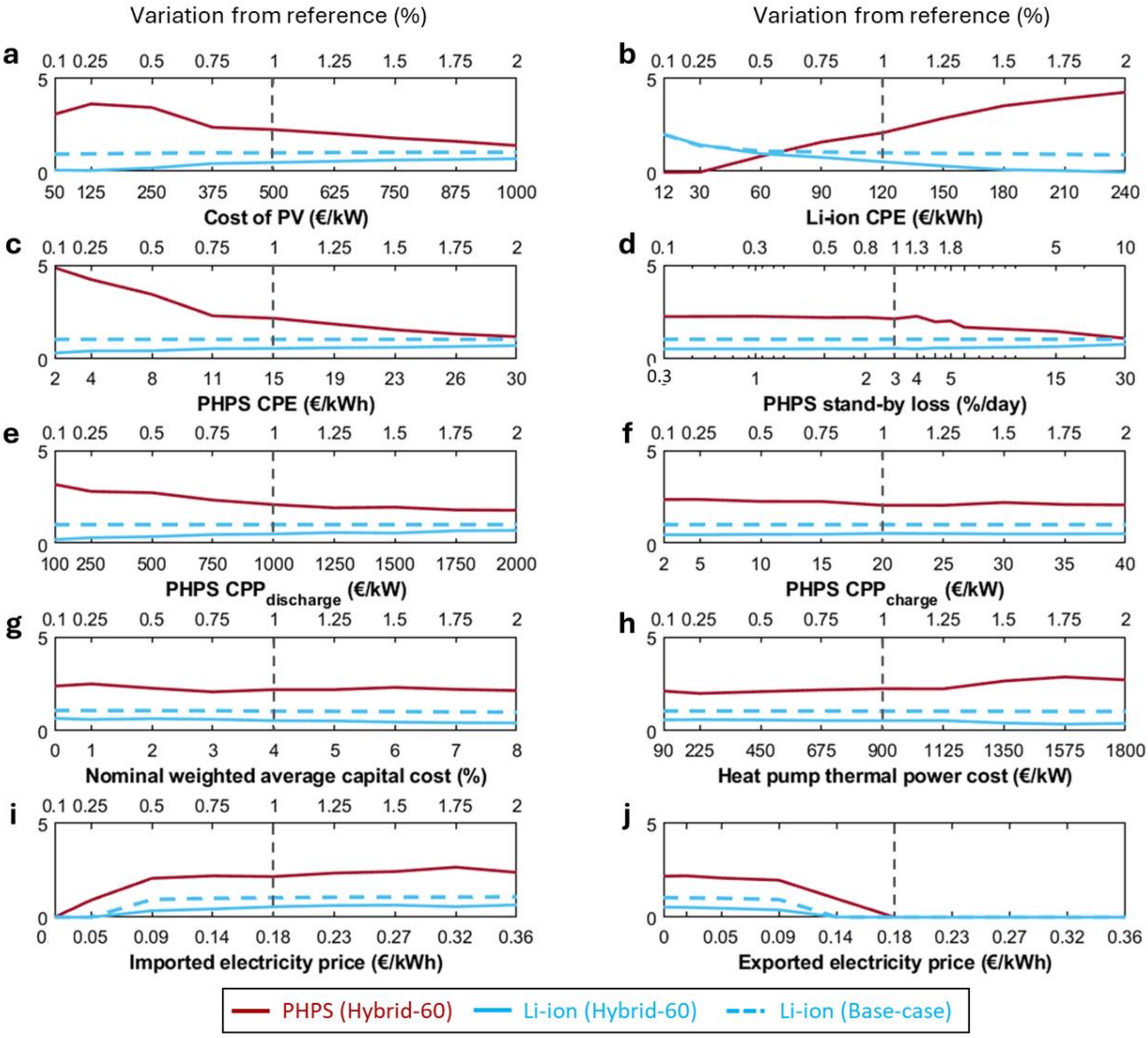

Figure 6. Optimal energy storage capacity (MWh) of PHPS (red lines) and Li-ion (blue lines) for the Hybrid-60 (solid line) and the base-case (dashed line) configurations. These results correspond to a sensitivity analysis of the cost of PV (panel a), Li-ion battery (panel b), CPE of PHPS (panel c) and its daily stand-by losses (panel d), $CPP_{PHPS\text{-}dis}$ (panel e) and $CPP_{PHPS\text{-}ch}$ (panel f) of PHPS, the nominal weighted average capital cost (panel g), heat pump's thermal power cost (panel h) and the cost of the imported (panel i) and exported (panel j) electricity from the grid. These results are shown for the block of apartments located in Madrid scenario, assuming COP=3, H2P efficiency=30 %.

## 3.4. Limitations and future work

These results may be subject to change due to the approximations and assumptions inherent in the simulated scenarios. Firstly, the model operates under a quasi-stationary regime, meaning that transient behaviours of all components are neglected. Converters and storage systems include some operation losses independent on the time-step or the state of charge. For instance, the maximum discharge power of the PHPS may depend on its state of charge, or the heat pump's COP should ideally incorporate flow velocities in its model. Additionally, losses associated with certain components, such as the water flow in the hot water tanks, heat exchangers and the heating demand side, have been neglected.

Madrid and Berlin were selected to represent warm and cold regions, respectively. To demonstrate the robustness of these findings, the analysis was extended to Los Angeles and Boston as analogous warm and cold regions in the United States (Appendix C) Figure C. 1 replicates the analysis shown in Figure 2 for Los Angeles instead of Madrid, and Figure C. 2 reproduces the analysis of Figure 5 for Los Angeles and Boston instead of Madrid and Berlin. Although the energy demands are comparable for the United States and European cities (see Table C. 1 ), differences among results arise primarily from the higher solar resource availability in Los Angeles and Boston relative to Madrid and Berlin. Despite these differences, the tendency and variations in both locations share similarities supporting the robustness and generalizability of the study's conclusions.

This study adopts an approach in which a consistent building demand scenario is first defined, and the relative performance improvement of incorporating PHPS technology is then assessed. Consequently, other potential improvements arising from the integration of energy-efficient or decarbonization measures in buildings, such as passive radiative cooling [64] or enhanced thermal insulation [65], are not considered. These measures, by reducing space heating demand, may impact on the feasibility of PHPS due to the resulting decrease in waste heat demand.

Furthermore, PHPS has been used as a Combined Heat and Power system, generating electricity through its Heat-to-Power converter, and utilizing the waste heat to supply the heating demand. An interesting direction for future research is to explore applications that directly supply the high-temperature stored heat to decarbonize hard-to-abate sectors.

Determining the time resolution is also crucial in this kind of dynamic modelling as a course resolution may overlook certain details in the operation. However, employing a smaller time-step is beyond the scope of this study due to the associated high computational cost. A 1-h time step has been adopted to capture key system dynamics, such as intra-day variations in PV generation and energy demand, while enabling an evaluation of the benefits of incorporating or not an energy storage system. Still, hourly resolution may disregard high power peaks occurring for limited durations, or changes in PV generation, which could potentially lead to under sizing the installation.

Other existing models, such as the one described in [66], or the software PyPSA [67], optimize energy dispatchability within the system. In particular, PyPSA incorporates a look-ahead optimization approach that utilizes forecasts of future generation and demand. In contrast, the model presented in this article manages energy flows based on causal decisions, meaning that decisions are made independently of the time step and without predicting future events. This approach is similar to the model already used in [28], [36], [49]. In this regard, although a comparison of different charging and discharging priorities has been conducted, a fixed decision tree could be a limitation and should be considered when interpreting the results.

Additionally, the price of the electricity imported from the grid is set exogenously and remains constant throughout the year, unlike other energy models such as PyPSA. Given the uncertainty associated with current electricity cost fluctuations, alternative scheduling strategies or electricity tariffs could be analysed to assess their impact. Furthermore, the model could be extended to include additional features, such as allowing energy storage systems to be charged using electricity imported from the grid when the electricity is cheap.

This modelling approach focuses on a building powered exclusively by PV generation and grid electricity, without considering the broader energy landscape. Further studies should expand this scope to analyse PHPS integration within larger energy systems, such as a district heating networks. These networks are commonly installed in northern countries [68], where heating demand is high. In these regions, long-duration energy storage is critical to address the intermittency of wind generation, which is more prevalent than solar. PHPS could play a key role here, particularly when deployed as a CHP system at scale, potentially reducing energy costs while balancing grid fluctuations.

While long duration energy storage is in principle better used for overcoming wind generation fluctuations [4]. This study has shown PHPS to be feasible when combined with solar PV systems. Thereby, it would also be valuable to assess the potential in southern regions, where PV and Li-ion batteries are considered as the most cost-effective solution [4]. Although district heating networks are currently less developed in these areas, some studies [69] advocate for its adoption as part of decarbonization strategies. Consequently, the role of PHPS in future energy systems may depend not only on techno-economic feasibility, but also on political decisions driving the transition to low-carbon and electricity networks.

# 4. Conclusions

This study evaluated the impact of utilizing waste heat generated during the discharge of a PHPS system, when operated in conjunction with Li-ion batteries and a photovoltaic (PV) installation for building self-consumption. A model was developed to simulate the energy supply of the building, with the objective of minimizing the levelized cost of energy for an optimum sizing of the main components.

The two different approaches for leveraging the waste heat generated during the heat-to-power conversion (i.e., discharging) of PHPS have been analysed. The first one (Hybrid-60) involves directly supplying the waste heat at the demand setpoint temperature of 60ºC, while the second approach (Hybrid-T) supplies the waste heat at a temperature, *T*, below the setpoint and uses it to enhance the COP of the heat pump. The benefits of leveraging the waste heat have been evaluated, focusing on how some key parameters, such as the heat-to-power conversion efficiency, the COP of the heat pump and the supply waste heat temperature, impact on PHPS feasibility.

The results gathered from this study demonstrate that the PHPS reaches greater feasibility when the waste heat is supplied directly at the demand setpoint temperature (Hybrid-60) rather than being used to boost the COP of the heat pump (Hybrid-T). However, in both cases, leveraging waste heat allows PHPS to become part of the optimal solution at lower H2P efficiencies. Both studied climates show advantages from this waste heat utilization approach, though hybridization yields greater LCOE reductions in southern regions due to the higher PV availability. These findings remain consistent across all four building types analysed.

The sensitivity analysis shows that these conclusions are subject to parameter assumptions, such as the cost per energy of PHPS and Li-ion battery, or the cost of electricity exported and imported from the grid.

# 5. Glossary

| **Nomenclature** | **Meaning** |
|---|---|
| COP | Coefficient of performance |
| CHP | Combined heat and power |
| CPE | Cost per energy capacity |
| $CPP_{HP}$ | Heat pump cost per power capacity |
| $CPP_{PHPS\text{-}dis}$ | Cost per discharge power capacity of PHPS |
| H2P | Heat to power |
| HTES | High temperature energy storage |
| LCOE | Levelized cost of energy |
| Li-ion | Lithium ion |
| LTES | Low temperature energy storage |
| P2H | Power to heat |
| PHPS | Power-to-heat-to-power storage |
| PV | Photovoltaic |
| RTE | Round-trip conversion efficiency |
| TPV | Thermo-photovoltaic |

## 6. Data availability

Data will be available on request.

## 7. Acknowledgements

This work has been funded by the European Union's Horizon Europe Research and Innovation Programme under grant agreement No 101057954, and by the European Union's Horizon 2020 Research and Innovation Programme "SDGine for Healthy People and Cities" in UPM under the Marie Sklodowska-Curie agreement No. 945139. Views and opinions expressed are however those of the author(s) only and do not necessarily reflect those of the European Union or European Innovation Council. Neither the European Union nor the granting authority can be held responsible for them.

## 8. Author contributions

Conceptualization, A.D.; data curation: A.L.C funding acquisition, A.D; investigation, A.L.C., A.D.; methodology, A.L.C., A.D.; resources, A.D; software: A.L.C, visualization: A.L.C., A.D.; writing – original draft: A.L.C., A.D.; writing – review & editing: A.L.C., A.D.

# Appendices

## A. System model and algorithm description

### A.1. System description and energy-decision making

An energy decision making tree (illustrated in Figure A. 1) manages the adequateness performance of the system, determining the priority orders of the energy flows in the system to ultimately cover the building energy demand. These decisions are independent on the time horizon, and do not rely on energy price or PV generation forecasting. The electricity generated by the PV installation and the electrical grid is used to: cover the electricity and heating demand, which includes the space heating and the domestic hot water; or charge the energy storage units.

The developed model simulates the energy demand supply (electricity and heat) of a building, assuming a fully electrified installation. This means that both electricity and heating demand are ultimately supplied by electrical sources. PV generation surpluses can be stored in the electrical storage systems (PHPS or Li-ion battery) and in the LTES, being used as electrical and heating back-ups respectively. From now on, only the Hybrid-60 and the Hybrid-T configuration will be explained, noticing that the base-case configuration is a special case of the hybrid configurations when PHPS is not present. The electrical storage systems are modelled as simplified containers, independently on the energy storage system type (i.e., thermal or electrochemical), which store and supply electricity with certain charging ($\eta_{charge}$) and discharging efficiency ($\eta_{disch}$). The discharging losses ($1-\eta_{disch}$) of PHPS are assumed to be the waste heat at a certain temperature. In the case of the Hybrid-60 configuration, the waste heat is directly supplied at 60 ºC and stored in the LTES. In the Hybrid-T configuration, the waste heat is supplied at a lower temperature and is stored in an additional waste heat water tank to store to be eventually delivered to enhance the COP of the heat pump. Both LTES are modelled as a container filled in with water at a certain temperature, which have self-discharging losses represented as a drop in the stored water temperature.

A summary of the priorities on the energy use are gathered in Table A. 1 for the Hybrid-60 and Hybrid-T configurations, also applicable for the base-case excluding PHPS. Electricity demand is covered preferably by the PV generation, then by the PHPS system, followed by the Li-ion battery, and finally by the electrical grid. Heating demand will be preferentially covered by the heat stored in the hot water tank, then by electricity from PV, PHPS, Li-ion and the grid to run both the heat pump and the heat resistance. Surplus of PV generation will be preferably used to charge the Li-ion battery, then the PHPS, and finally the LTES. If there is still PV generation excess, it will be injected to the grid.

It has been established that Li-ion battery is charged before PHPS as its high CPE and conversion efficiency makes the energy stored more valuable than the one stored in the PHPS. As mentioned in the introduction, Li-ion optimum energy capacity is low and its discharge power is high, being better used to supply high power peaks. On the contrary, PHPS has high cost per discharge power ($CPP_{PHPS\text{-}dis}$) and low CPE, outcoming as a storage with a large energy capacity and relatively low output power, being suitable for base-load operations. Therefore, PHPS is settled in the decision-making tree to be discharged first at its output power to save Li-ion when a power demand peak occurs. In other words, Li-ion battery is discharged only when the PHPS is fully discharged or if the power demand exceeds the maximum discharging power capacity of the PHPS (typically low). This approach takes advantage of the low cost per power (CPP) of Li-ion battery and the low CPE of PHPS. Despite the energy flow priority order shown in Table A. 1 may influence on the optimum sizing of the components, studying the optimum existing (dis)charging strategies is out of the scope of this study.

Table A. 1. Energy flow priority order.

| Electricity demand | Heat demand | Storage of surplus generation |
|---|---|---|
| 1. PV | 1. Heat from LTES | 1. Li-ion |
| 2. PHPS | 2. Electricity from PV | 2. PHPS |
| 3. Li-ion | 3. Electricity from PHPS | 3. LTES |
| 4. Grid | 4. Electricity from Li-ion | |
| | 5. Electricity from Grid | |

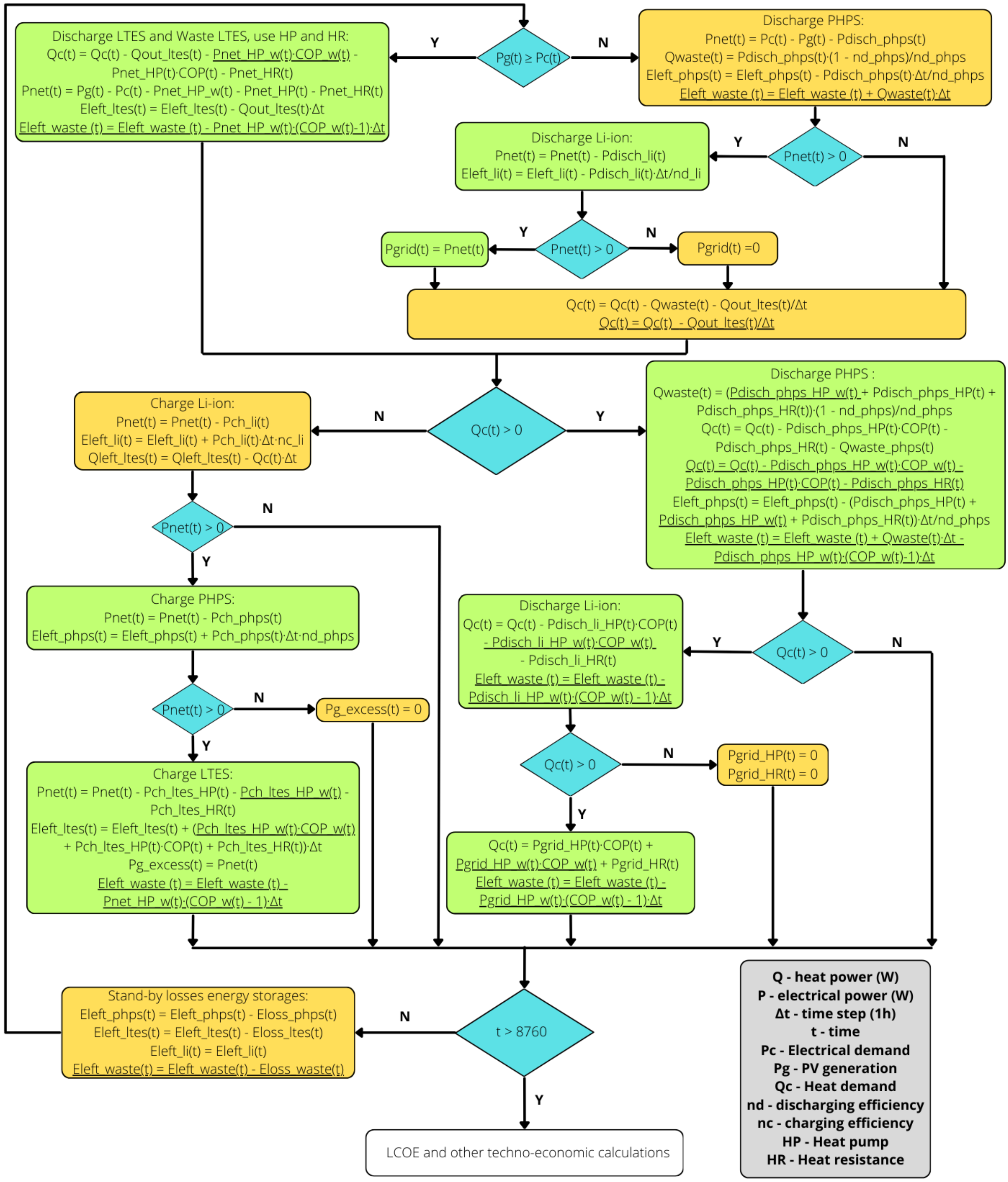

Figure A. 1. Energy decision making tree for the Hybrid-60 configuration. The operations are applicable to the Hybrid-T configuration are underlined.

### A.3. Energy demand and PV generation

Data from PV generation has been simulated in PVSyst for Madrid and Berlin, assuming optimum tilt for each location [54] and standard PV modules and inverters. The PV generation corresponds to the hourly output data from simulations of a 10-kW installation. Then, our model scales up or down the PV generation with the optimal PV power capacity obtained from the optimization program. The authors are aware that simplifying the PV generation by scaling the original 10 kW installation does not consider optimal design of a PV installation (such as number of parallel or series PV modules, number of inverters, etc.) or its exact losses, but assessing the precise impact of these effects is out of the scope of the study. The ambient temperature used to run the simulations in PVSyst is assumed to be the typical meteorological year retrieved from EnergyPlus [70]for that location (illustrated in panels a and c in Figure A. 2 for Madrid and Berlin). This temperature data coincides with the one used to calculate the buildings' energy demand in Energy Plus. Therefore, energy demand and PV generation are correlated with the same weather conditions. Results from PVSyst for the 10-kW installation normalized to 1 kW are shown in panels b and d in Figure A. 2 for Madrid and Berlin. As a reference, the PV generation of an installation of 10 kW is 17 MWh/y for Madrid and 11 MWh/y for Berlin.

The input data used by Energy Plus to calculate the energy demand of a certain building and location are the previously mentioned ambient temperature, and the specific building design (e.g., occupation schedule, building materials, etc.). The data of the buildings is obtained from the buildings' examples found in [71]. Normalized hourly heating and electricity demand resulting for a block of apartments, a hospital, a hotel and an office located in Madrid are depicted in Figure A. 3 below. The same data is depicted for Berlin in Figure A. 4.

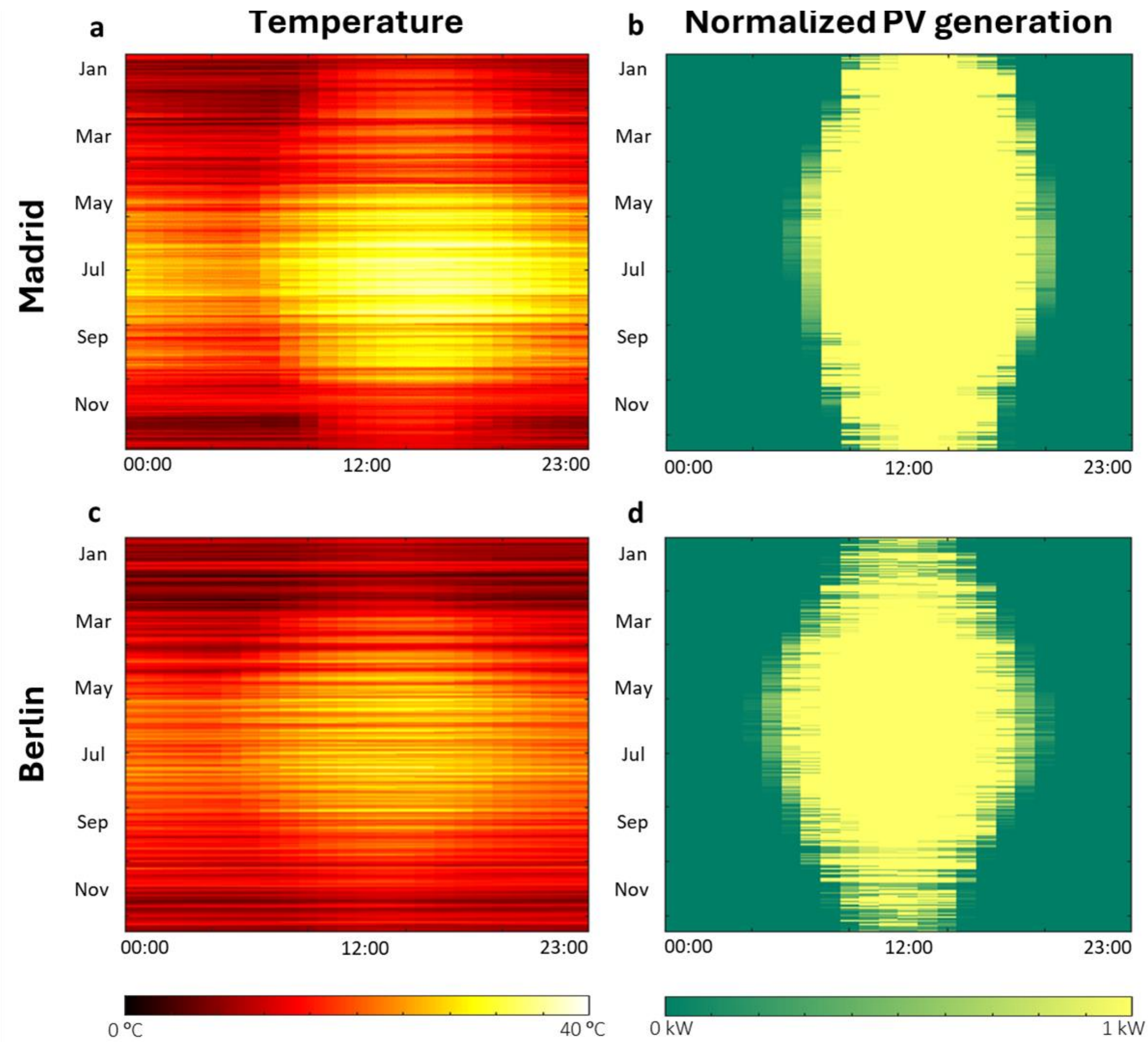

Figure A. 2. Panels a and c depict the ambient temperature, and panels b and d the normalized PV generation for each hour (x-axis) and day (y-axis) of the year. These results correspond to the location of Madrid (panels a and b) and Berlin (panels c and d).

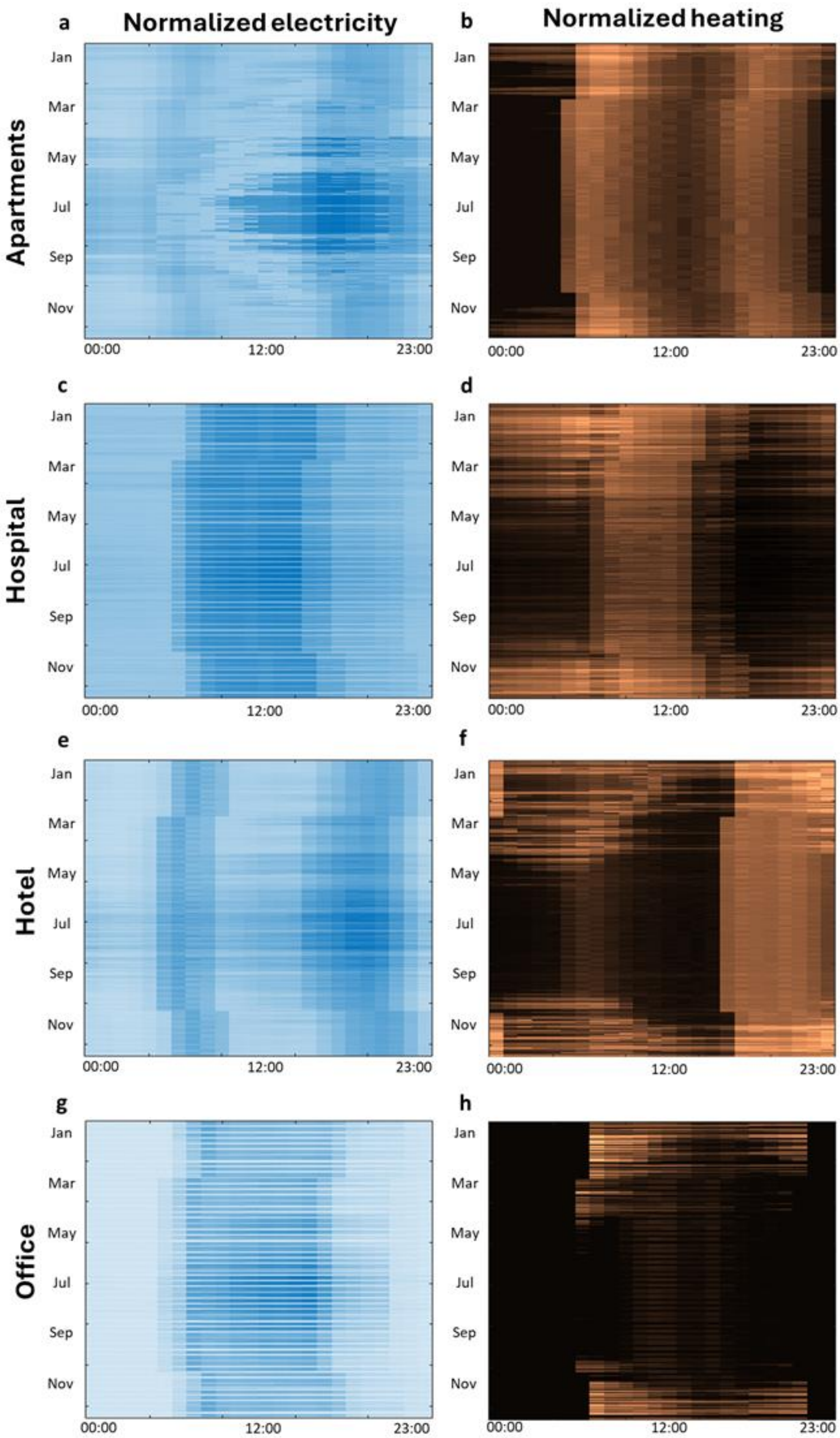

Figure A. 3. Electricity and heating demand normalized to their respective maximum electrical or heating power for each hour (x-axis) and day (y-axis) of the year. These results were obtained from simulating a block of apartments (panels a and b), a hospital (panels c and d), a hotel (panels e and f) and an office (panels g and h) in Energy Plus. Buildings are located in Madrid.

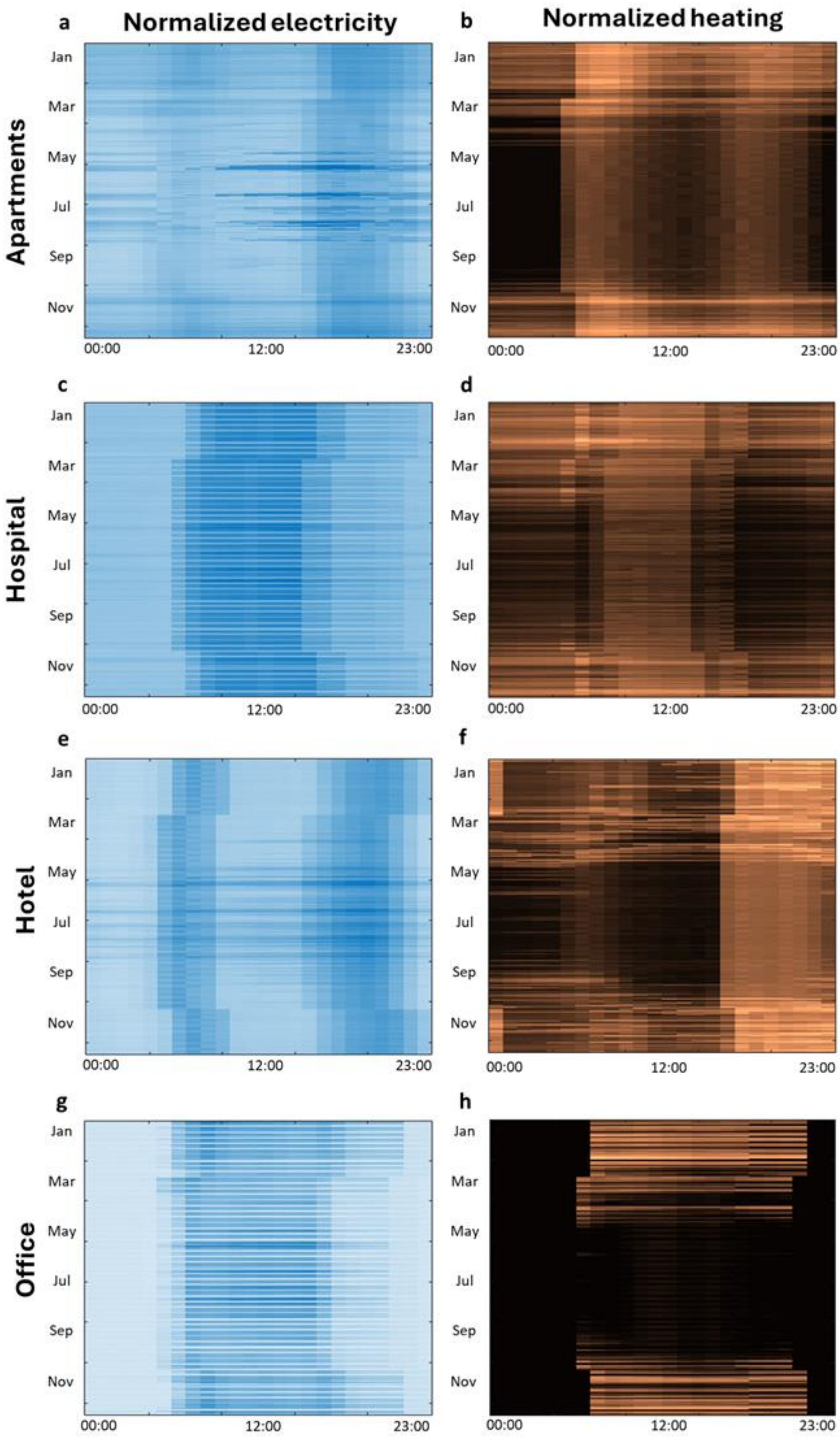

Figure A. 4. Electricity and heating demand of a building normalized to their respective maximum electrical or heating power for each hour (x-axis) and day (y-axis) of the year. These results were obtained from simulating a block of apartments (panels a and b), a hospital (panels c and d), a hotel (panels e and f) and an office (panels g and h) in Energy Plus. Buildings are located in Berlin.

A.4. Coefficient of Performance (COP) calculation

Similar to the definition by Rana et al. [7], the model considers the heat pump to be a black box with certain conversion efficiency from electrical to heating energy, defined as COP. The COP in the model has been calculated from the Carnot efficiency equation considering the heat pump to be an ideal heat engine in reverse operation. In order to apply certain operating losses, the ideal COP of the heat pump has been multiplied by an efficiency value as seen in Eq. (A.1) below:

$$COP_{HP} = \eta_{HP} \cdot \frac{T_H}{T_H - T_C} = \frac{Q_{demand}}{W} = \frac{Q_C + W}{W} \tag{A.1}$$

Where the $COP_{HP}$ depends on the heat pump's efficiency $\eta_{HP}$ (%), the hot side temperature $T_H$ (K) and the cold side temperature $T_C$ (K). The hot temperature refers to the supply temperature (in this case, it is the demand setpoint temperature 60 ºC or 333.15 K). The cold temperature indicates the temperature of the heat source (e.g., ambient temperature or waste heat temperature). The COP is also calculated through the total heat demand, $Q_{demand}$ (J), or the heat pumped from the cold side, $Q_C$ (J), and the $W$ (J), which is the electricity required to pump the heat from the cold to the hot side.

Given the uncertainty of the efficiency value ($\eta_{HP}$), the COP has been calculated as a function of the cold temperature ($T_C$) for a range of $\eta_{HP}$ values, and plotted in graphs in Figure A. 5. These graphs have been compared to other COP equations found in literature: COP values from Staffel et al [52] and Kozarcanin et al. [62] have been depicted in the same graph in solid and dashed black lines respectively. Instead of using these in-built approximations found from literature, Eq. (A.1) is selected for considering the COP tendency towards infinity when the difference of the hot and cold temperature approaches to zero. Thus, when the waste heat temperature approaches the setpoint temperature, the electricity required to provide heat is significantly reduced. The minimum operating temperature ($T_C$) of the heat pump is -10 ºC, aligned with values found in literature [12], [62], [63], [72], [73].

Given the closer alignment of the graph assuming $\eta_{HP}$=35% (yellow line) with COP values from literature for typical ambient temperatures ranging from 0 to 30 ºC, our model adopts this value for subsequent COP calculations.

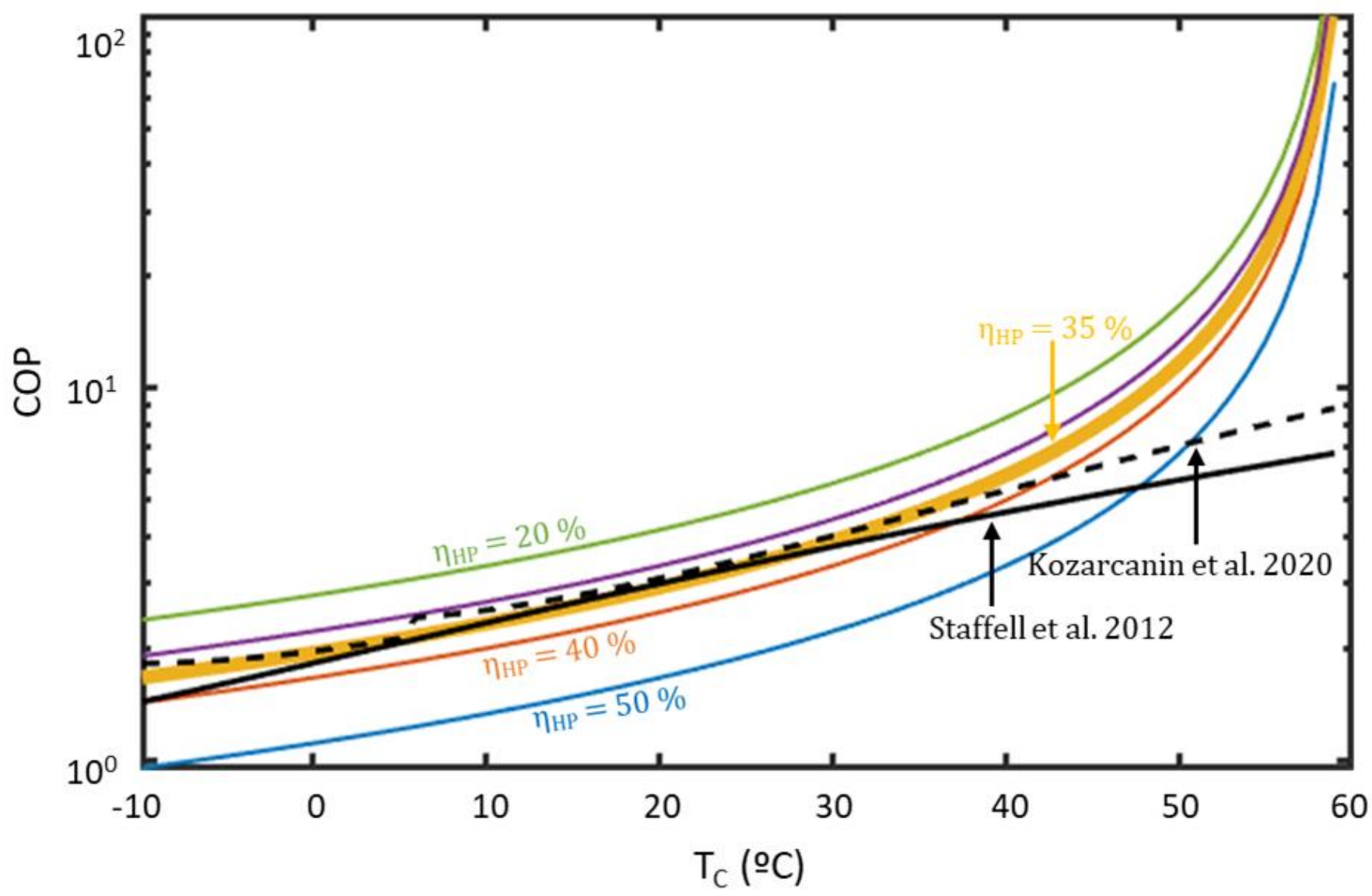

Figure A. 5. COP calculation following Eq. (A.1) as a function of the cold temperature ($T_C$) and assuming $T_H$=60 ºC for different values of the heat pump's efficiency (coloured lines). Black (solid and dashed lines) show the COP following the equations in Staffell et al.[52] and Kozarcanin [62] et al. respectively. The model assumes a heat pump's efficiency of 35%, highlighted by a yellow line.

The developed model described in this appendix calculates the COP based on the availability of waste heat stored in the waste heat water tank, and the ambient and waste heat temperatures. If the waste heat water tank is empty or the waste heat temperature is below the ambient temperature, the COP is calculated considering $T_C$ as the ambient temperature. Otherwise, the COP is calculated based on the waste heat temperature. When using the waste heat, $Q_C$ is the energy stored in the hot water tank, which is modelled as a container, being charged or discharged by adding or removing energy. Then, the electricity required ($W$ in Eq. (A.1)) to meet the heating demand is calculated resolving the same equation knowing the heating demand ($Q_{demand}$) and the COP calculated from the cold and hot temperatures.

In order to validate Eq. (A.1), the hourly COP has been calculated over a year assuming the ambient temperature ($T_C$) to be the typical meteorological year found in [71] and $T_H$=60 ºC. The resulting average COP is 2.7 for Madrid, and 2.4 for Berlin, aligning with values reported in Kozarcanin [62]: ~2.6 in Madrid and ~2.3 in Berlin with historical ambient temperature data.

The authors are aware that this model is a simplification from reality as it neglects certain physical phenomena such as the flow, or the heat transferred through the exchangers. However, a more detailed model of this system is out of the scope of our study.

A.5. Optimization model

After feeding this model with its inputs: heating and electricity demand, PV generation, and economic and technical assumptions, the simulation is run with hourly resolution for a year following the already described system control. The resulting yearly data is used to calculate the levelized cost of energy (LCOE) defined in Eq. (1) which is minimized by the optimum sizing of the main components in the system. The optimization algorithm is based on the Nelder-Mead method [74] and is employed with a set of initial seed values to mitigate the risk of local minima. The optimized parameters, indicated in Table 1, are the energy and charging/discharging power storage capacities (for PHPS and Li-ion batteries); PV power capacity; hot water tank energy capacity; the heat pump power capacity; and if applicable, the waste heat water tank energy capacity. Optimum solutions of the Hybrid-60 and Hybrid-T configurations may result in either a combination of both batteries (i.e., PHPS and Li-ion) or in a single battery solution.

The LCOE is calculated from the capital expenditure $CAPEX$ and $OPEX$, shown in equations (A.2) and (A.3), divided by the total energy demand.

$$\begin{aligned} CAPEX = {} & C_{PHPS} + C_{li-ion} + C_{rep-Li-ion} + C_{LTES} + C_{PV} + C_{HP} = CPE_{PHPS} \cdot E_{cap-PHPS} \\ & + CPP_{in-PHPS} \cdot P_{in-PHPS} + CPP_{out-PHPS} \cdot P_{PHPS} + CPE_{li-ion} \cdot E_{cap-li-ion} \\ & + CPP_{in} \cdot P_{in-li-ion} \\ & + \sum_{i=1}^{N_{repl}} \frac{1}{(1+WACC_{real})^{i*L_{Li-ion}}} (CPE_{li-ion} \cdot E_{cap-li-ion} + CPP_{in} \\ & \cdot P_{in-li-ion}) + CPP_{PV} \cdot P_{PV} + CPE_{HW} \cdot E_{cap-LTES} + CPE_{HW} \\ & \cdot E_{cap-waste-LTES} + CPP_{HR} \cdot P_{cap-HR} + CPP_{HP} \cdot P_{HP} + CPE_{HW} \\ & \cdot E_{cap-waste-LTES} \end{aligned} \quad (A.2)$$

$$\begin{aligned} OPEX = {} & (1+inf)^t \cdot (W_{grid} \cdot C_{var-grid} + P_{max-grid} \cdot C_{fix-grid} - W_{exp-grid} \\ & \cdot C_{exp-var-grid}) \end{aligned} \quad (A.3)$$

The $CAPEX$ (in €, equation A.2) depends on the cost of the electrical storage, $C_{PHPS}$ and $C_{li-ion}$, defined by their cost per energy capacity, $CPE_{PHPS}$ and $CPE_{li-ion}$ (€/kWh), their cost per input and output power capacity, $CPP_{in-PHPS}$, $CPP_{in-li-ion}$and $CPP_{out-PHPS}$ (€/kW$_{el}$) respectively, multiplied by the corresponding storage's capacities: maximum energy capacity, $E_{cap-PHPS}$ and $E_{cap-li-ion}$ (kWh) and maximum input and output power: $P_{in-PHPS}$ $P_{in-li-ion}$, $P_{out-PHPS}$ (kW$_{el}$). Li-ion's replacement is defined as $C_{rep-Li-ion}$, which is calculated for every required replacement ($N_{repl}$) over the installation lifetime. All the components within the installation are assumed to last during the installation lifetime (25 years, Table 1), except for Li-ion battery. Its replacement cost depends on the real weighted average capital cost ($WACC_{real}$), defined as per Eq. (A.4) below; and the Li-ion's lifetime ($L_{Li-ion}$). The LTES cost, $C_{LTES}$, which is its cost per energy and power capacities, $CPE_{LTES}$ (€/kWh) and $CPP_{HR}$ (€/kW), multiplied by the hot water tank and waste heat water tank energy and heat resistance power capacities, $E_{cap-LTES}$ (kWh), $E_{cap-waste-LTES}$ (kWh) and $P_{cap-HR}$ (kW). The PV installation cost, $C_{PV}$, which depends on the cost per power capacity, $CPP_{PV}$ (€/kW), and the total power installed, $P_{nom-PV}$ (kW). And finally, the heat pump cost $C_{HP}$, calculated multiplying the specific cost $CPP_{HP}$ (€/kW$_{el}$) by the maximum electrical power, $P_{HP}$ (kW$_{el}$).

The $OPEX$ depends on the energy bought from the grid, $W_{grid}$ (kWh), multiplied by the cost in that time period, $C_{var-grid}$ (€/kWh); the maximum contracted power, $P_{max-grid}$ (kW) multiplied by the cost per power installed in that period, $C_{fix-grid}$ (€/kW); minus the electricity exported to the grid, $W_{exp-grid}$ (kWh), multiplied by its cost, $C_{exp-var-grid}$ (€/kWh). In addition, the maintenance cost is included depending on its specific power cost, $C_{maintenance}$ (€/kW) and the nominal PV power installed, $P_{PV}$ (kW).

$$WACC_{real} = \frac{1 + WACC_{nom}}{1 + inf} - 1 \tag{A.4}$$

The $WACC_{real}$, referred to the discount rate, is calculated(A.4) from the inflation ($inf$) and the nominal weighted average capital cost ($WACC_{nom}$), defined in Table 1.

## B. Further results

B.1. Results for a block of apartments, a hospital, a hotel and an office for the scenarios in section 3.1

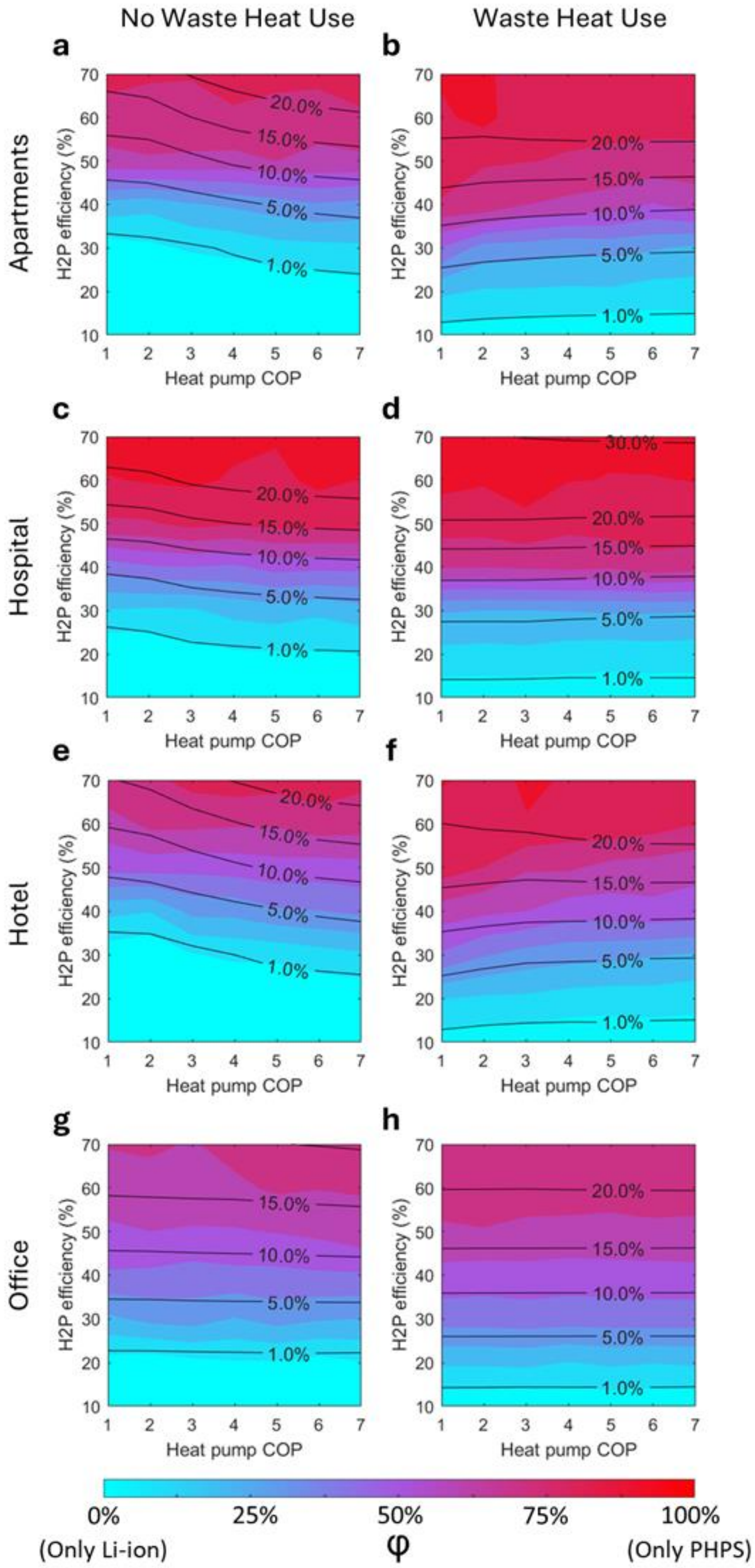

Figure B. 1. Results for a block of apartments (panels a and b), a hospital (panels c and d), a hotel (panels e and f), and an office (panels g and h) located in Madrid with variations in the COP of the heat pump (x-axis) and the H2P efficiency of the PHPS (y-axis). Panels a, c, e and g show

scenarios in which the waste heat is not leveraged, while panels b, d, f and h illustrate scenarios in which the waste heat is utilized to cover heating demand. Black lines depict the LCOE reduction (%) for the Hybrid-60 configuration compared to the base-line scenario (only Li-ion battery). Coloured contour plots represent the percentage contribution, in the total capital expenditures, of each energy storage system within the optimal hybrid solution, defined as $\varphi = \mathrm{CAPEX_{PHPS}}/(\mathrm{CAPEX_{PHPS}} + CAPEX_{Li-ion}) \cdot 100$, which is the share of PHPS in the total investment in electrical storage: 100% indicates a configuration relying entirely on PHPS (red-coloured region), whereas 0% represents a system using only Li-ion batteries (blue-coloured region).

B.2. Optimal capacities of Hybrid-60 configuration in the block of apartments from section 3.1

Figure B. 2 includes the optimal solution assuming COP=3 varying the H2P efficiency for the block of apartments located in Madrid. Results include PV capacity (panel a), and the discharging (panel b) and charging capacities (panel c) of the PHPS and Li-ion battery. Apart from its high energy capacity, PHPS provides the system with decoupled charging and discharging power capacities, unlike Li-ion. The discharging and charging capacities of Li-ion (blue lines) are identical for using the same energy converter in both operations. This results in a distinct threshold between the optimal charging and discharging capacities of Li-ion battery. The optimal system is designed to set its maximum discharging power to match the peak energy demand (0.2 MW), as indicated by the solid black line in panel b of Figure B. 2. Consequently, its charging capacity is constrained, as well as its ability to balance high PV peaks effectively.

In contrast, the discharging power capacity of PHPS is limited by its high cost, aligning its optimal capacity to the average demand power (depicted by the black dashed line in panel b) and resulting into large discharge ratios. Nevertheless, its low-cost favours large charging power and energy capacities, rendering PHPS an effective storage solution for mitigating PV variability and achieving high self-consumption ratios.

As an example, Figure B. 3 depicts the payback period for the Hybrid-60 configuration in the case of not using the waste heat (panel a) and using the waste heat (panel b), against the Base-case configuration payback period (panel c).

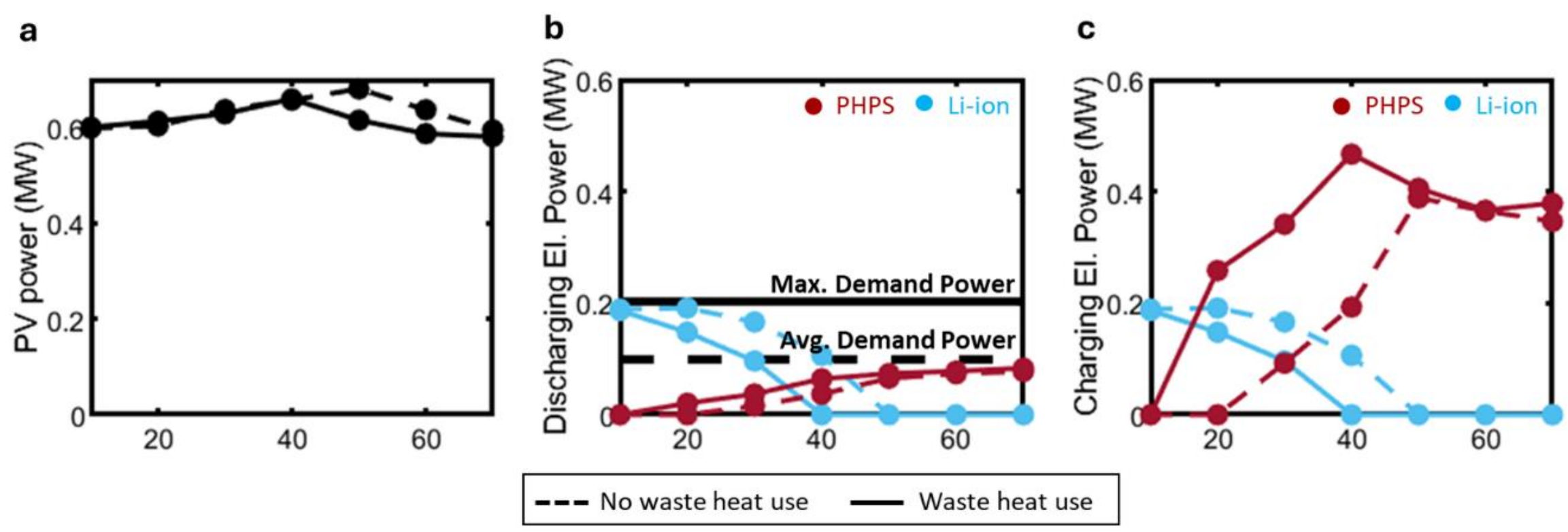

Figure B. 2. Results of the optimal solution for the block of apartments building when COP=3 and varying the H2P efficiency not using (dashed lines) and using (solid lines) the waste heat. These panels depict: (a) optimal PV power (MW), (b) discharging and (c) charging optimal electrical power capacities (MW). Panels b and c depict information for the PHPS (red) and Li-ion batteries (blue) in the optimal solution. Panel b includes the maximum and average demand power (black solid and dashed lines).

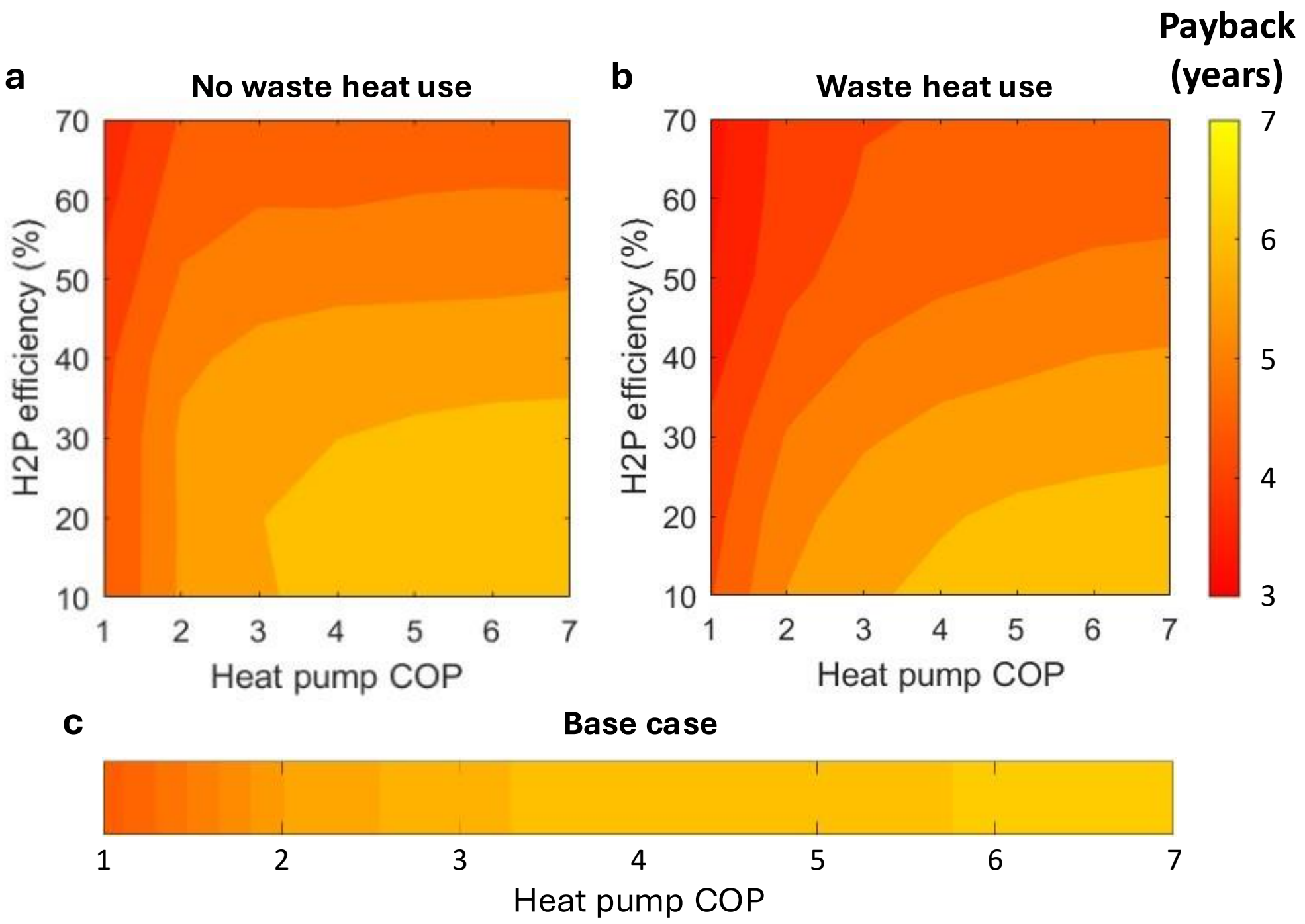

Figure B. 3. Payback period (years) for the Hybrid-60 configuration in a block of apartments located in Madrid with variations in the COP of the heat pump (x-axis) and the H2P efficiency of the PHPS (y-axis). Panel a shows a scenario in which the waste heat is not leveraged, while panel b illustrates a scenario in which the waste heat is utilized to cover heating demand. Panel c below depicts the payback period for the Base-case configuration.

B.3. Results for a block of apartments, a hospital, a hotel and an office for the scenarios in section 3.2

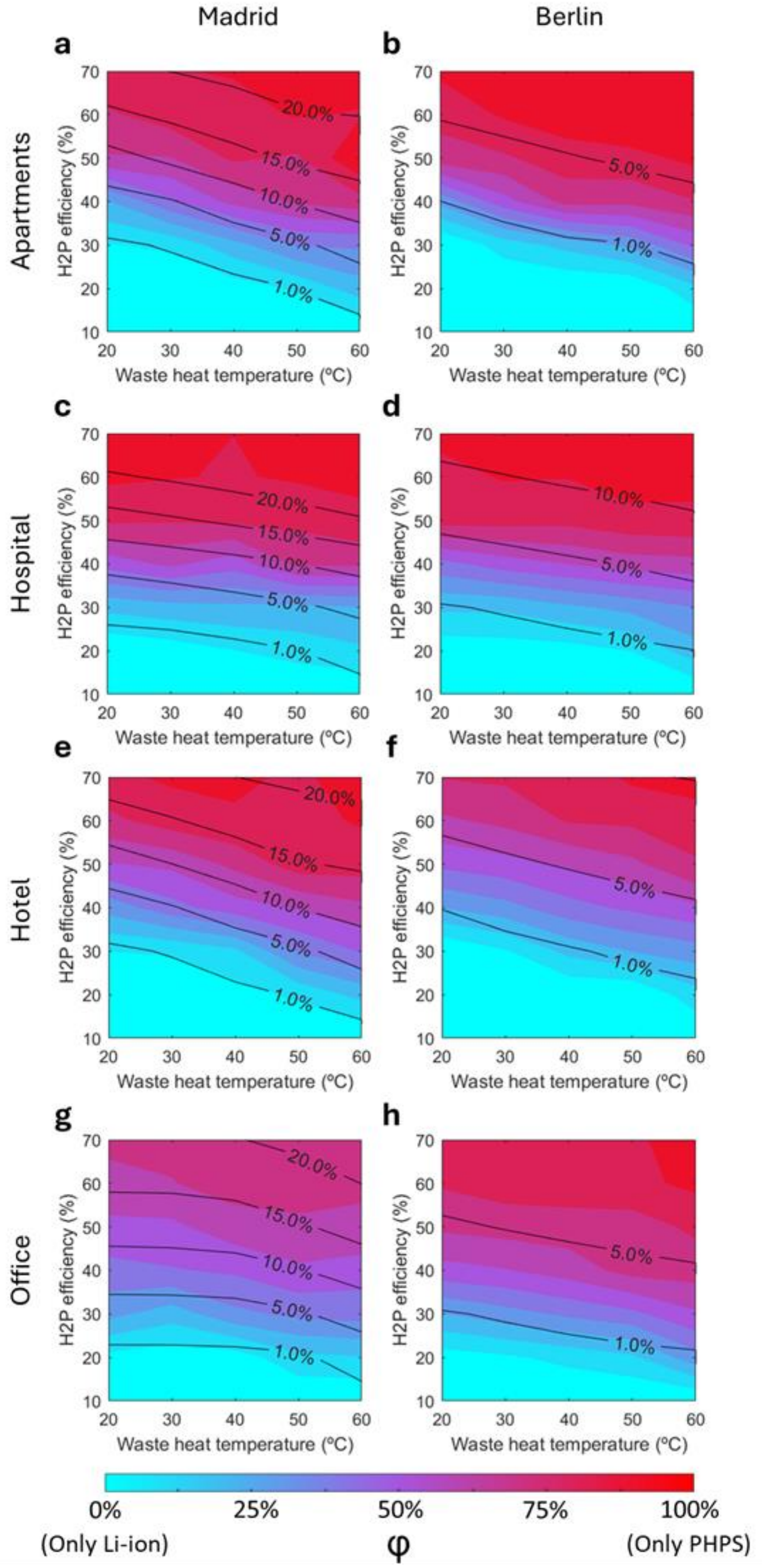

Figure B. 4. Results for a block of apartments (panels a and b), a hospital (panels c and d), a hotel (panels e and f), and an office (panels g and h) with variations in the waste heat temperature (x-axis) and the H2P efficiency of the PHPS (y-axis). Panels a, c, e and g show scenarios located in Madrid, while panels b, d, f and h illustrate scenarios located in Berlin. Black lines depict the LCOE reduction (%) for the Hybrid-T configuration compared to the base-line scenario (only Li-ion battery). Coloured contour plots represent the percentage contribution, in the total capital expenditures, of each energy storage system within the optimal hybrid solution, defined as $\varphi = \mathrm{CAPEX_{PHPS}}/(\mathrm{CAPEX_{PHPS}} + CAPEX_{Li-ion}) \cdot 100$, which is the share of PHPS in the total investment in electrical storage: 100% indicates a configuration relying entirely on PHPS (red-coloured region), whereas 0% represents a system using only Li-ion batteries (blue-coloured region).

B.4. Optimal capacities of Hybrid-T configuration in the block of apartments from section 3.2

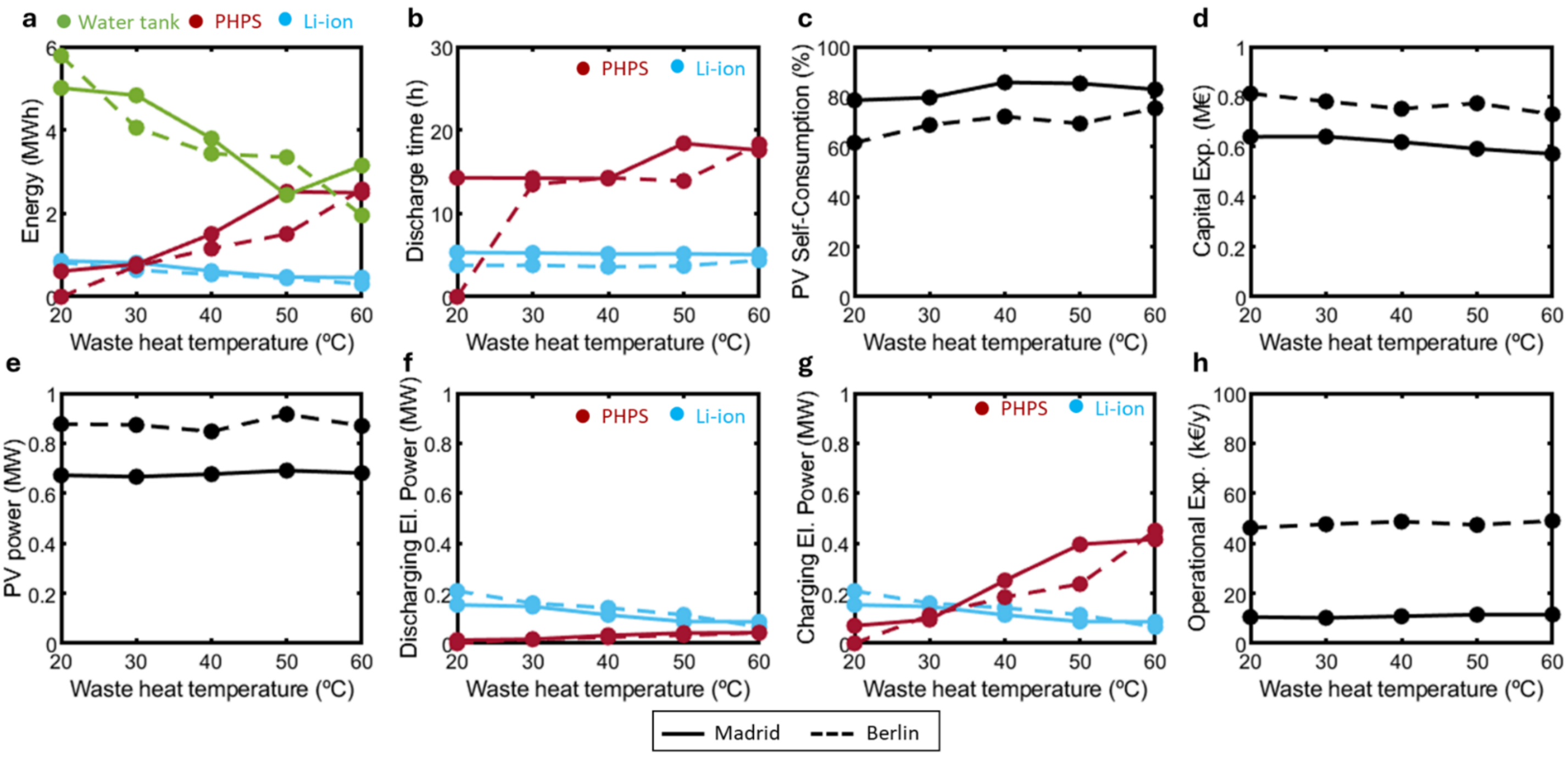

Figure B. 5. Results of the optimal solution for the block of apartments building when H2P efficiency of PHPS is 30%, varying the waste heat supply temperature in Madrid (solid lines) and Berlin (dashed lines). Panels depict: (a) optimal energy capacity of energy storage components (MWh), (b) discharge time and (c) PV self-consumption ratio (%), (d) system capital expenditure or CAPEX (M€), (e) optimal PV power, (f) optimal discharging and (g) charging electrical power capacities (MW) and (d) yearly operational expenditure or OPEX (k€/y). Panels a, b, f and g depict information for the PHPS (dark red), Li-ion batteries (blue) and water tanks (green), defined as the addition of both water tanks energy capacity.

## Results for Los Angeles and Boston

Table C. 1. Information for each type of building and location. Heat to energy demand ratio is defined as the heat to total energy demand.

| Location | Los Angeles | | | Boston | | |
|---|---|---|---|---|---|---|
| Yearly PV generation 1 kW installation | 1.9 MWh | | | 1.6 MWh | | |
| Building \ Parameter | Electricity demand (MWh/y) | Heating demand (MWh/y) | Heat to total energy demand (%) | Electricity demand (MWh/y) | Heating demand (MWh/y) | Heat to total energy demand (%) |
| Block of apartments | 523 | 270 | 34 | 580 | 400 | 41 |

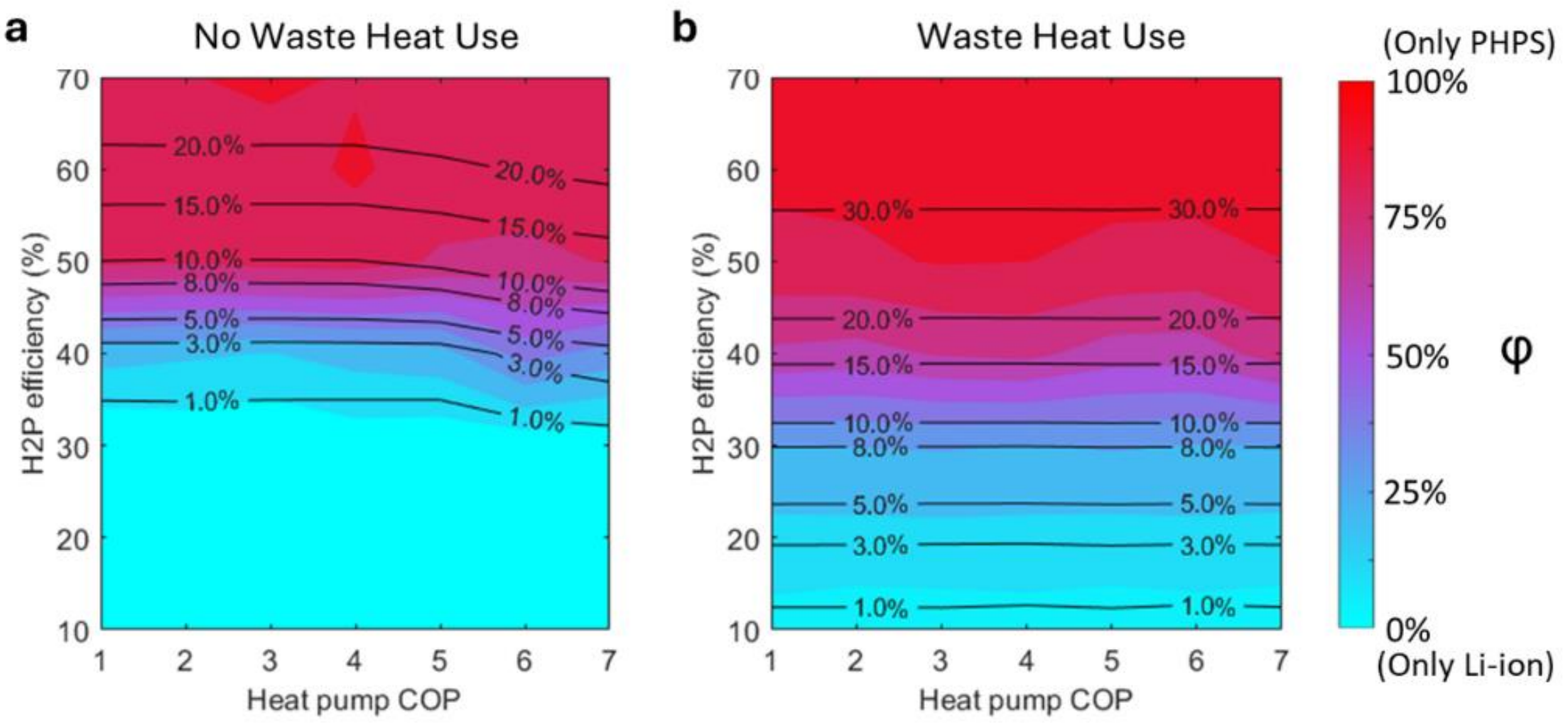

Figure C. 1. LCOE reduction (black solid lines) for the Hybrid-60 configuration compared to the base-line scenario (only Li-ion battery) for a block of apartments located in Los Angeles with variations in the COP of the heat pump (x-axis) and the H2P efficiency of the PHPS (y-axis). Panel a shows a scenario in which the waste heat is not leveraged, while panel b illustrates a scenario in which the waste heat is utilized to cover heating demand. Contour plot shows the percentage contribution of each energy storage system within the optimal hybrid configuration defined as $\varphi = \mathrm{CAPEX_{PHPS}}/(\mathrm{CAPEX_{PHPS}} + CAPEX_{Li-ion}) \cdot 100$, which is expressed as the share of PHPS in the total investment in electrical storage: 100% indicates a configuration relying entirely on PHPS (red-coloured region), whereas 0% represents a system using only Li-ion batteries (blue-coloured region).

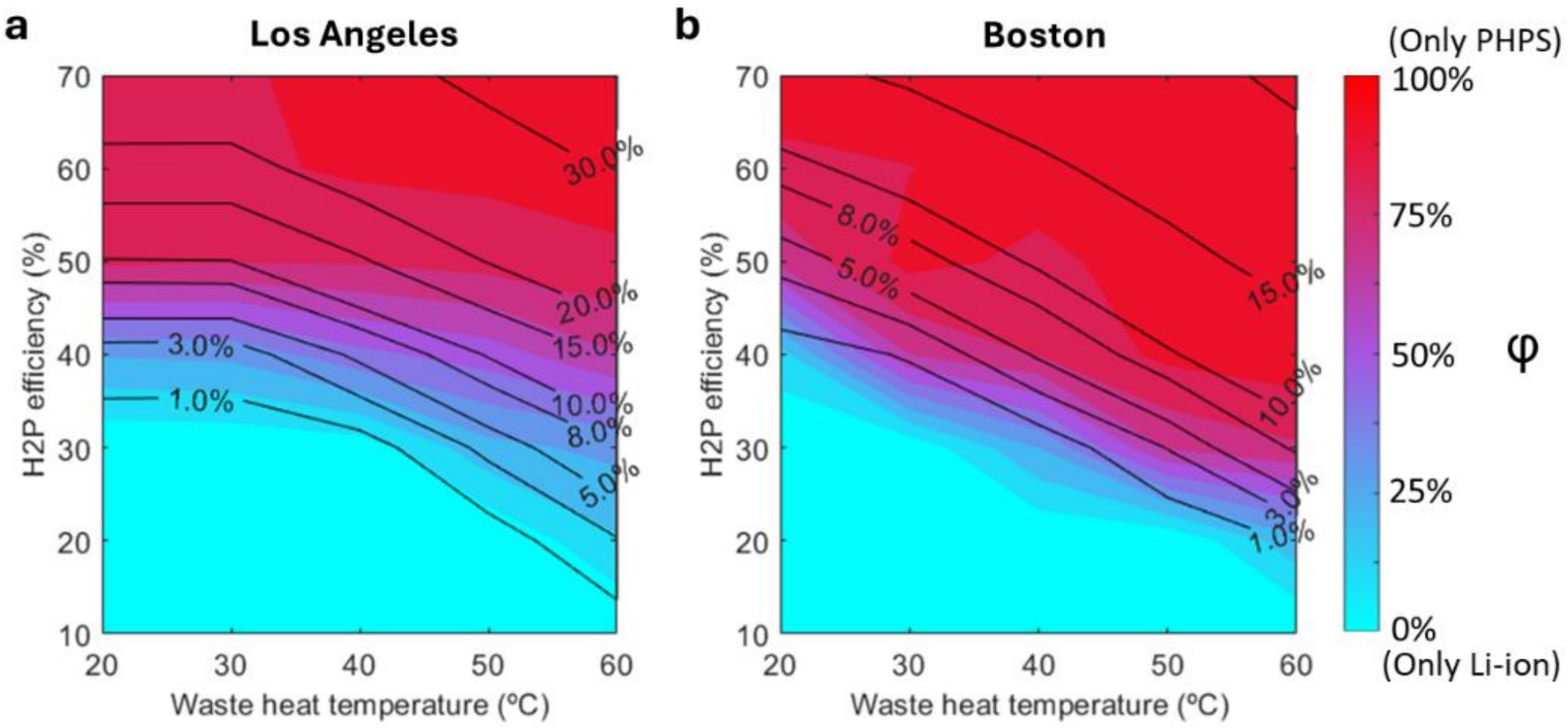

Figure C. 2. Reduction in the LCOE (solid lines) of the Hybrid-T configuration relative to the base-case configuration for a block of apartments in Los Angeles (panel a) and Boston (panel b). Results are provided as a function of the waste heat temperature and the H2P conversion efficiency of the PHPS. Coloured contour plots represent the percentage contribution, in the total capital expenditures, of each energy storage system within the optimal hybrid solution, defined as $\varphi = \mathrm{CAPEX}_{\mathrm{PHPS}}/(\mathrm{CAPEX}_{\mathrm{PHPS}} + CAPEX_{Li-ion}) \cdot 100$, which is the share of PHPS in the total investment in electrical storage: 100% indicates a configuration relying entirely on PHPS (red-coloured region), whereas 0% represents a system using only Li-ion batteries (blue-coloured region).